\documentclass{aa}  

\usepackage{graphicx}
\usepackage[colorlinks=true, citecolor=blue, linkcolor=blue, breaklinks=true]{hyperref}
\makeatletter
\renewcommand*\aa@pageof{, page \thepage{} of \pageref*{LastPage}}
\makeatother
\usepackage{natbib}

\usepackage{txfonts}
\usepackage{bm}
\begin{document}

   \title{On the tomographic cluster clustering as a cosmological probe}


   \author{M. Romanello
          \inst{1,2}
          \and
          F. Marulli\inst{1,2,3}
          \and
          L. Moscardini\inst{1,2,3}
          \and
          C. Giocoli\inst{2,3}
          \and
          G. F. Lesci\inst{1,2}
          \and
          S. Contarini\inst{1,4}
          \and
          A. Fumagalli\inst{5,6,7}
          \and
          B. Sartoris\inst{5,7} 
          }

   \institute{Dipartimento di Fisica e Astronomia “A. Righi” - Alma Mater Studiorum Università di Bologna, via Piero Gobetti 93/2, 40129 Bologna, Italy
   \and
   INAF - Osservatorio di Astrofisica e Scienza dello Spazio di Bologna, via Piero Gobetti 93/3, 40129 Bologna, Italy 
   \and 
   INFN - Sezione di Bologna, Viale Berti Pichat 6/2, 40127 Bologna, Italy
   \and 
   Max Planck Institute for Extraterrestrial Physics, Giessenbachstr. 1, 85748 Garching, Germany
   \and
   Universitäts-Sternwarte München, Fakultät für Physik, Ludwig Maximilians-Universität München, Scheinerstrasse 1, 81679, München, Germany
   \and
   IFPU, Institute for Fundamental Physics of the Universe, via Beirut 2, 34151 Trieste, Italy
    \and
   INAF-Osservatorio Astronomico di Trieste, Via G. B. Tiepolo 11, 34143 Trieste, Italy
\\\\
        \email{massimilia.romanell2@unibo.it}
             }

   \date{}

 
  \abstract
  {The spatial distribution of galaxy clusters is a valuable probe for inferring fundamental cosmological parameters. We measured the clustering properties of dark matter haloes from the \textsc{Pinocchio} simulations, in the redshift range $0.2 < z < 1.0$ and with virial masses $M_\mathrm{vir} > 10^{14} M_\odot \, h^{-1}$, which reproduce the expected mass selection of galaxy cluster samples. The past-light cones we analysed have an angular size of 60 degrees, which approximately corresponds to a quarter of the sky. We adopted a linear power spectrum model, accounting for nonlinear corrections at the baryon acoustic oscillations scale, to perform a comparative study between 3D and 2D tomographic clustering. For this purpose, we modelled the multipoles of the 3D two-point correlation function, $\xi(r)$, the angular correlation function, $w(\theta)$, and the angular power spectrum, $C_\ell$. We considered observational effects such as redshift-space distortions, produced by the peculiar velocities of tracers, and redshift errors. We found that photometric redshift errors have a more severe consequence on the 3D than on the 2D clustering, as they affect only the radial separation between haloes and not the angular one, with a relevant impact on the 3D multipoles. Using a Bayesian analysis, we explored the posterior distributions of the considered probes with different tomographic strategies, in the $\Omega_m-\sigma_8$ plane, focusing on the summary parameter $S_8\equiv \sigma_8\sqrt{\Omega_m/0.3}$. Our results show that in the presence of large photometric errors the 2D clustering can provide competitive cosmological constraints with respect to the full 3D clustering statistics, and can be successfully applied to analyse the galaxy cluster catalogues from the ongoing and forthcoming Stage-III and Stage-IV photometric redshift surveys.}


   \maketitle
%
\section{Introduction}
The growth of cosmic structures is governed by long-range gravitational interactions between a large number of systems, which behave as tracers of the total matter distribution. Among them, according to the hierarchical model for the cosmic structure formation, galaxy clusters are the latest objects produced by the gravitational infall and the merging of dark matter haloes \citep{Bardeen86, Tormen98}. Reaching masses of up to $10^{15}M_\odot$ and radii of the order of a few megaparsec \citep{McNamara07}, galaxy clusters are the largest gravitationally bound systems ever formed, located at the nodes of the filamentary cosmic web of large-scale structures (LSS). Such astrophysical objects are multi-component systems, made mainly of dark matter and, to a lesser extent, of baryonic matter in different phases \citep{Voit05, Allen11}. This, together with the variety and number of physical processes involved, including both the nonlinear and non-local character of the gravitational force, makes a fully analytical treatment of their formation and evolution not accurate enough, considering the precision of current and future datasets. \\
\indent For this reason, numerical simulations of the LSS are widely used. Since the most relevant effect in the evolution of large-scale density fluctuations is the gravitational interaction, and since the dominant component is the collisionless dark matter, which does not involve computationally expensive hydrodynamical effects, observations of galaxy clusters can be reproduced, within certain limits, by $N$-body simulations floowing only the dark matter component. In fact, the baryonic component has only a negligible effect on the formation and evolution of galaxy clusters, and hence on in their clustering properties \citep{Springel18, Hernandez23}. In particular, $N$-body simulations rely on a fiducial cosmology, defined by a certain set of cosmological parameters, and evolve the positions of particles undergoing gravitational interaction over cosmic time \citep{Springel01, Springel05}. Therefore they require a considerable computational effort to realise a sufficient number of mocks necessary for an adequate statistical treatment of the results. \\
\indent Despite advances in computational techniques in recent years, there are still some serious limitations to the generation of cosmological simulations \citep{Garaldi20, Vogelsberger20, Hernandez23}. An accurate description of the evolution of dark matter perturbations is fundamental for modelling the clustering properties of galaxy clusters. First of all, the study of clustering requires an accurate sampling of the baryon acoustic oscillations (BAO) scale, at around $100$ Mpc $h^{-1}$, which is only possible for large-size simulations \citep{Monaco16}. Typical requirements for such mock catalogues imply a volume of about one Gpc$^3$ and a mass resolution below $10^{10} \, M_\odot \, h^{-1}$, and become even more prohibitive if we want to produce a wide past-light cone (PLC), without multiple replications along the same line of sight \citep{Monaco13, Monaco16}. Furthermore, a large number of simulations corresponding to $N$ independent realisations of the Universe are needed to evaluate the associated errors and covariance matrices \citep{Manera13, Fumagalli_counts, Fumagalli_clustering}. In fact, a smaller number of realisations would make the estimate very noisy \citep{Giocoli17, Giocoli20}, with relevant consequences for the cosmological constraints, also because of the problems related to the inversion of the covariance matrix \citep{Manera13, Munari17}. Thus, approximate methods are often used to obtain independent realisations in a much faster way, by
evolving the density fluctuations with the linear theory, and computing higher-order corrections through a perturbative approach \citep[see e.g.][for a review]{Monaco16}.\\
\indent Despite the fact that cluster catalogues usually contain sparse statistics, they can be used extensively in cosmological analyses. In particular, cluster clustering has a strong dependence on the amplitude of the mass power spectrum and the matter density parameter of the Universe \citep[see e.g.][]{Sartoris16}, so it can be successfully exploited to constrain these cosmological parameters \citep{Veropalumbo14, Sereno15, Veropalumbo16, Marulli18, Lindholm21, Marulli21, Lesci22, Fumagalli24_SDSS, Romanello24}. It can also significantly improve the constraining power of other cosmological probes, such as such as weak gravitational lensing of galaxy clusters \citep{Sereno15} and number counts \citep{Sartoris16}, when analysed in combination with them. \\
\indent Currently, cluster clustering is less studied than the traditional galaxy clustering; nevertheless it has a series of remarkable advantages. Being hosted by the most massive virialised haloes, galaxy clusters are more clustered than galaxies, i.e. they are more biased tracers \citep[e.g.][]{Mo96, Moscardini01, Sheth01, Hutsi10, Allen11}. In contrast to the galaxy bias, which is usually considered to be a nuisance parameter in cosmological analysis because difficult to be correctly modeled, especially at small comoving separation, the effective bias of galaxy cluster samples can be predicted theoretically \citep{Branchini17, Paech17, Lesci22}, by means of fitting functions that depend on the mass and redshift of the host haloes \citep[e.g.][]{Tinker10}. Furthermore, the BAO peak in the correlation function of galaxy clusters exhibits a low nonlinear damping and is sharper compared to other tracers, i.e. closer to the predictions of linear theory \citep{Veropalumbo16}. Finally, galaxy clusters present smaller peculiar velocities, thus the effect of redshift-space distorsion (RSDs) is reduced, allowing us to simplify the theoretical model for the clustering signal. \\
\indent Clustering information is often investigated with the three-dimensional (3D) two-point correlation function, $\xi(r)$, measured from the 3D comoving spatial coordinates of the tracers, which however are not directly accessible, requiring a cosmological assumption to convert the observed redshifts and angular positions to distances. Since the fiducial cosmology assumed for the measurements might be different from the true one, the measured two-point correlation function might be geometrically distorted by the so-called \textit{Alcock-Paczynski} (AP) effect \citep{Alcock1979}. A way to avoid this is to extract the clustering signal from the angular positions alone, using the two-point angular correlation function, $w(\theta)$, in configuration space, and its harmonic-space counterpart, the angular power spectrum, $C_\ell$ \citep[][]{Hauser73, Peebles73}, which do not require any AP correction \citep{Asorey12, Salazar14, Camera18}. In principle, these two statistics carry the same cosmological information when the entire spectrum of spherical harmonics and angular scales is considered. However, in practice the limited range of wavelengths that can be probed, in the reduced volume of real catalogues, affects the correlation function and the power spectrum differently \citep{Ata18, Avila18}. Furthermore, we expect a different response to noise and to potential uncorrected observational or systematic effects, which may lead to complementary, though highly correlated, estimates of cosmological parameters \citep{Sanchez17, Camacho19, Romanello24}.\\
\indent In this paper, we make use of a set of dark matter-only simulations to investigate the clustering properties of massive haloes, which reproduce some observational conditions of Stage-III
surveys, such as the Kilo Degree Survey\footnote{\url{http://kids.strw.leidenuniv.nl/}} \citep[KiDS; see][]{deJong17, Kuijken19}, the Dark Energy Survey\footnote{\url{ https://www.darkenergysurvey.org}} \citep[DES; see][]{DES16}, the Hyper Suprime-Cam\footnote{\url{https://hsc.mtk.nao.ac.jp/ssp/}} (HSC) Subaru Strategic Program \citep[HSC-SSP; see][]{Aihara18}, and the number of clusters expected in Stage-IV surveys, e.g. \textit{Euclid}\footnote{\url{http://sci.esa.int/euclid/}} \citep{Laureijs11, Blanchard20, Scaramella22, Mellier24}. In particular, we aim at exploring the best strategies for maximising the cosmological information that can be extracted from tomographic angular cluster clustering, in both its variants, i.e. $w(\theta)$ and $C_\ell$, to verify if it can provide competitive cosmological constraints with respect to the full 3D study. This requires to manage with an appropriate binning strategy and to handle observational effects, such as RSDs and photometric errors, quantifying their impact on the modelling of the clustering signal.\\
\indent The paper is organised as follow: in Sect. \ref{sect_PINOCCHIO} we describe the dark matter-only simulations, and the adopted angular and mass selections. In Sects. \ref{sect_3D_correlation}, \ref{sect_angular_correlation_function} and \ref{Sect_angular_power_spectrum} we present the methods to measure and model $\xi(r)$, $w(\theta)$ and $C_\ell$. In Sect. \ref{sect_bayesian} we show the result of the Bayesian cosmological analysis. Finally, in Sect. \ref{sect_conclusions}, we draw our conclusions and present future perspective. In Appendix \ref{Appendix_A} we quantify the impact of reducing the sky coverage on the angular clustering, while in Appendix \ref{sect_model_covariance} we test some simple analytic models for the covariance matrices of the angular correlation function and power spectrum. \\
\indent The management of halo catalogues, the measurements of the relevant statistical quantities, the cosmological modelling and the Bayesian inference presented in this study are performed with \textsc{CosmoBolognaLib}\footnote{\url{https://gitlab.com/federicomarulli/CosmoBolognaLib}} \citep{MarulliCBL}, a set of \textit{free} software C++ and Python libraries.
\section{Data: PINOCCHIO simulations}
\label{sect_PINOCCHIO}
\indent \textsc{Pinocchio} \citep[PINpointing Orbit-Crossing Collapsed HIerarchical Objects algorithm;][]{Monaco02, Monaco13, Monaco16, Munari17}
is a fast approximated code for the production of dark matter halo catalogues, which works as follows. A Gaussian density field is generated on a cubic grid and then smoothed on a set of scales $R$. Following the extended Press \& Schechter approach \citep{Press74, Bond91}, the ellipsoidal collapse model is then used to estimate the collapse time for each particle on the grid, considering all the smoothing scales. The ellipsoid evolution is approximated by the third-order Lagrangian Perturbation Theory (LPT), which is valid until an orbit crossing event occurs or the ellipsoid collapses on its shortest axis. LPT is also used to calculate halo displacements \citep{Monaco16, Munari17}. When the collapse occurs, the particle is expected to be assigned and incorporated into dark matter haloes or in the filaments between them. This so-called \textit{fragmentation} is done according to the two main processes that characterise the hierarchical clustering of dark matter haloes, namely matter accretion and merging \citep{Monaco02, Monaco13, Munari17}.\\
\indent In this way, the code is able to reproduce the mass and the accretion histories of dark matter haloes \citep{Monaco16}, with an accuracy for the halo mass function and the linear halo bias of $5\%$ and $10\%$, respectively, compared to the full $N$-body simulations \citep{Fumagalli_counts, Fumagalli_covariance, Fumagalli_clustering}. Even an accuracy of 10\% in the power spectrum can be achieved up to $z \approx 1$. This means that the LPT is able to reproduce the clustering of haloes, placing them in the right position even at relatively low redshifts \citep{Monaco16}, with a big advantage in terms of computational time. For these reasons \textsc{Pinocchio} has been used for several \textit{Euclid} preparatory science papers \citep[see e.g.][]{Fumagalli_counts, Fumagalli_clustering, Keih22}.
\subsection{Dark matter halo catalogues: angular and redshift selections}
\label{sec_pinocchio_catalogue}
\indent The \textsc{Pinocchio} algorithm simulates cubic boxes with side of $L = 3870$ Mpc $h^{-1}$ and periodic boundary conditions, from which replication PLCs are produced, selecting only those dark matter haloes that are causally connected to an observer at the present time \citep{Fumagalli_counts, Fumagalli_clustering}. We make use of a set of 1000 PLCs, with an aperture of approximately $60$ deg, for a final effective area of $10\,313$ deg$^2$, i.e. a quarter of the sky. To study the clustering properties of haloes, we focus only on one mock, while the full set of PLCs is used to build the random catalogue (see Sect. \ref{sec_random}) and to evaluate the numerical covariance matrices. The simulations have been run assuming \citet{Planck14} cosmology: $\Omega_m = 0.30711$, $\Omega_b = 0.048254$, $h = 0.6777$, $n_s = 0.96$, $\sigma_8 = 0.8288$ and, unless otherwise stated, this is assumed to compute the clustering models that are compared with the measured quantities and presented in the next figures.\\
\indent The full catalogue contains haloes in the redshift range $0<z<2.5$, with virial masses $M_\mathrm{vir}>6.7 \times 10^{13} M_\odot \, h^{-1}$, sampled with more than 50 particles. However, we focus only on the interval $0.2<z<1.0$. In fact, in real catalogues, a lower limit of $0.1-0.2$ is expected, due to the reduced lensing power that does not allow a robust lensing analysis, necessary for the mass calibration and the following cosmological exploitation \citep{Bellagamba19, Giocoli21, Lesci24}. The upper limit is quite conservative and depends on several considerations. First, it is because our work is in preparation for the analysis of the cluster catalogues form the next KiDS data releases (KiDS-DR4, \citealp{Kuijken19}; KiDS-DR5, \citealp{Wright24}), where the expected number of clusters at $z>1$ is very low and certainly not sufficient for a mass calibration with weak lensing, as already verified in KiDS-DR3, which is extended only up to 0.6 \citep[see e.g.][]{Bellagamba19, Giocoli21, Lesci_counts, Lesci22, Giocoli24, Romanello24}. Furthermore, the low number of clusters at $z>1$ implies that the shot noise term dominates the signal even at the largest angular scales, with non-negligible consequences for the angular power spectrum measurements. Finally, at higher redshifts, we expect a general reduction in the completeness and purity of the cluster sample affecting the robustness of the cosmological analysis. \\
\indent As explained in \citet{Fumagalli_counts}, we use a version of the catalogue in which the halo masses have been rescaled in order to match the \citet{Despali16} halo mass function model, preserving all the fluctuations due to sample variance and shot noise. We also select only haloes with a virial mass larger than  $10^{14}$ $M_\odot \, h^{-1}$, which in good approximation corresponds to the predicted limits of the Stage-IV photometric survey cluster selection function, i.e. the limiting cluster mass threshold at $z<2$ \citep[e.g.][]{Sartoris16}. The PLCs are available in both real and redshift space. From the latter, we have also produced mock catalogues by introducing a Gaussian redshift perturbation at the redshift $z$ of each halo, which we will hereafter refer to as photo-$z$ space:
\begin{equation}
\label{eq_gaussian_perturbation_photoz}
    P(z_\mathrm{phot}|z)=\frac{1}{\sqrt{2\pi}\sigma_z} \mathrm{exp}\left[-\frac{\delta z^2}{2\sigma_z^2}\right],
\end{equation}
where $\delta z=z_\mathrm{phot}-z$, so that there is no systematic bias in the photometric distribution of the tracers. On the other hand the standard deviation follows the typical shape for photometric redshift surveys: 
\begin{equation}
\label{eq_sigmaz_1z_pinocchio}
    \sigma_z=\sigma_{z,0}(1+z).
\end{equation}
The choice of the $\sigma_{z,0}$ mimics the characteristics of the observations. In particular, we are interested in broad-band photometric data, for which we set $\sigma_{z,0}=0.05$, corresponding to the accuracy required for the Stage-IV galaxy photometric surveys \citep[see e.g.][]{Sartoris16, Adam19}. In general, the redshift errors associated with cluster detection are smaller than the corresponding errors in galaxy photometry, because they are computed using the information coming from a set of galaxies. For this reason, we should consider the previous value as a worst-case scenario and also adopt a value of $\sigma_{z,0}=0.02$, which corresponds to the photometric uncertainty of the KiDS-DR3 cluster catalogue \citep{Maturi19}. We divide our catalogue either in four redshift bins of binwidth 0.2, or two redshift bins of binwidth 0.4. We limit our analysis to broad redshift bins, several times larger than the photometric errors, while we do not evaluate the narrow bin case, where $\Delta z \lesssim \sigma_z$, because the redshift subsampling dramatically increases the relative importance of the shot noise over the signal, given the limited number of massive haloes. The chosen binwidths give us a minimum of $\Delta z \approx 2\sigma_z$ (for $\Delta z=0.2$, $\sigma_{z,0}=0.05$ and in $0.8<z<1.0$) and a maximum of $\Delta z\approx 14 \sigma_z$ (for $\Delta z=0.4$, $\sigma_{z,0}=0.02$ and in $0.2<z<0.6$). 
\subsection{Random catalogue}
\label{sec_random}
The construction of the random catalogue is fundamental to avoid biases in the measurement of the 3D and angular correlation functions. The random catalogue must take into account all the possible selection effects associated with the construction of the data catalogue, e.g. irregularities in the mask due to satellite tracks, stars, or the Milky Way plane. In this case, such effects are limited to the angular and redshift distribution of dark matter haloes within the footprint mask of the simulation. So we build our random catalogue by randomly extracting cluster positions and redshifts from the full set of 1000 halo mocks. This process
is repeated for real, redshift and photo-$z$ space. In this way the random light cone reproduces the R.A., Dec and redshift average distributions of \textsc{Pinocchio}, which are independent of the individual realisations. To limit the shot noise effects, the size of the random catalogue is 10 times larger than the halo one.
\section{3D two-point correlation function}
\label{sect_3D_correlation}
\subsection{Characterisation of the measurement and covariance matrix}
The 3D two-point correlation function is defined as the excess probability of finding a pair of haloes in the comoving volume elements $\delta V_1$ and $\delta V_2$, at the comoving distance $r$, relative to that expected from a random distribution. It is expressed by:
\begin{equation}
\label{eq_dP_12}
\delta P_{12}(r) = n_V^2\,[1+\xi(r)]\,\delta V_1 \delta V_2\,,
\end{equation}
where $n_V$ is the mean number density of objects per unit comoving volume. The two-point correlation function can be measured with the \citet{Landy93} (LS) estimator:
\begin{equation}
\label{eq_LandySzalay}
    \xi_{\mathrm{LS}}(r)=\frac{DD(r)+RR(r)-2DR(r)}{RR(r)},
\end{equation}
where $DD(r)$, $RR(r)$ and $DR(r)$ are the number of data-data, random-random and data-random pairs with separation $r\pm\Delta r/2$, respectively, normalised for the total number of data-data, random–random and data-random pairs. Eq. \eqref{eq_LandySzalay} provides an unbiased measurement of the two-point correlation function for a number of random objects $N_R \rightarrow \infty$.
\begin{figure*}[htbp]
\centering
\includegraphics[width=1\textwidth]{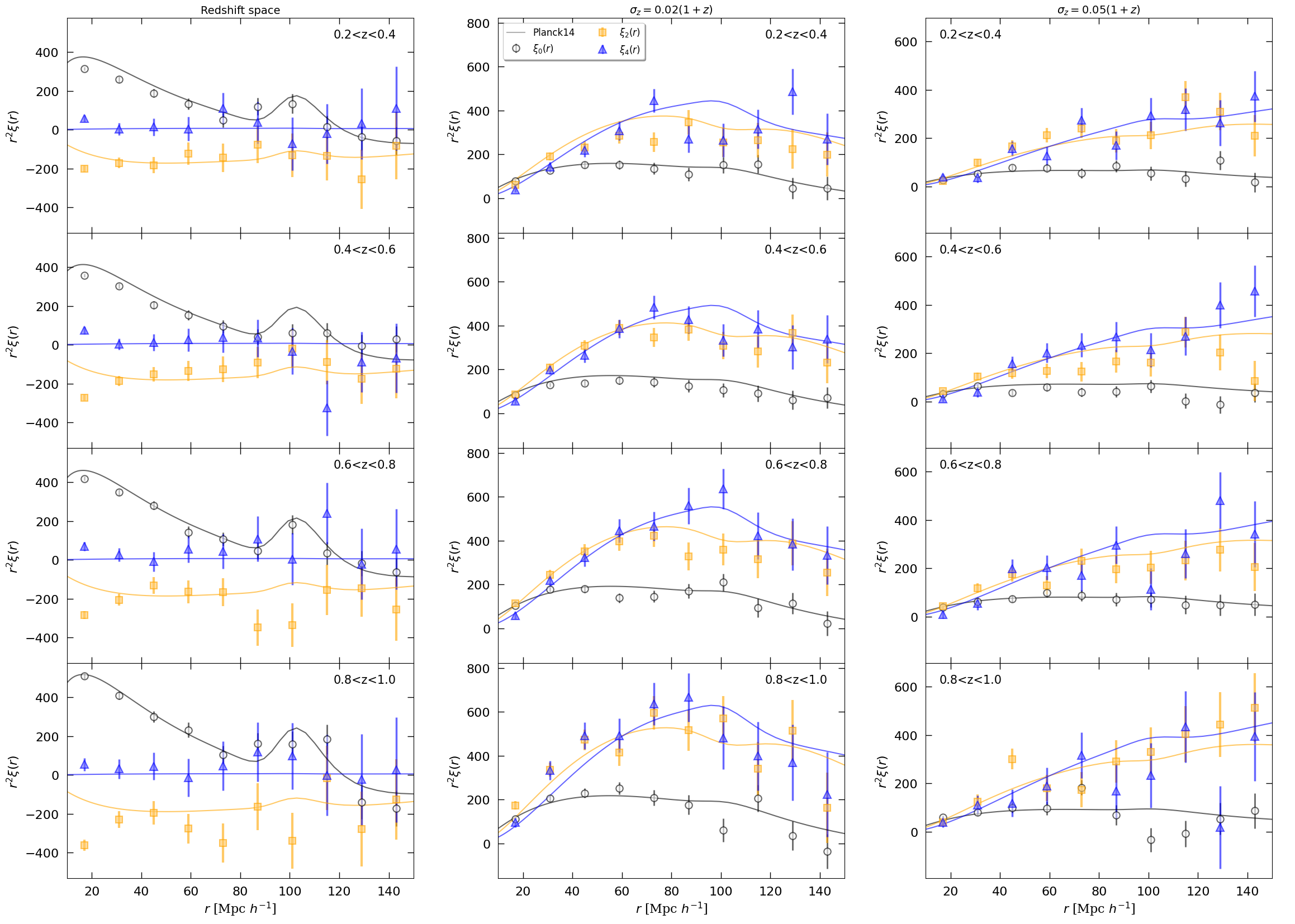}
\caption{The 3D two-point correlation function first three even multipoles, measured from one \textsc{Pinocchio} PLC. Errors are computed as the square root of the diagonal elements of the covariance matrix. We show results for monopole (white circles), quadrupole (orange squares), and hexadecapole (blue triangles). Different columns refer to redshift space, with $\sigma_z=0$, and photo-$z$ space, with $\sigma_z=0.02(1+z)$ and $\sigma_z=0.05(1+z)$. The solid lines show the theoretical two-point correlation function multipoles, computed from the power spectrum estimated with CAMB, assuming a linear model with $\Sigma_\mathrm{NL}=0$, i.e. without nonlinear damping effects at the BAO scale. They include systematic effects, such as RSDs and photometric damping. The colour of the lines corresponds to the multipoles indicated by the symbols.}
\label{fig_3D_correlation_pinocchio_02}
\end{figure*}
In this work, the 3D two-point correlation is measured in 10 linearly equispaced radial bins, in the range $15-150$ Mpc $h^{-1}$. At these scales, the BAO peak is probed and the halo bias can be approximated as a constant \citep{Manera11, Mandelbaum13, Fumagalli_clustering}, i.e. it does not depend on $r$, but only on redshift and mass. The first three even multipoles of the 3D correlation function, $\xi_0(r)$, $\xi_2(r)$ and $\xi_4(r)$, are shown in Fig. \ref{fig_3D_correlation_pinocchio_02} for all the redshift bins of our analysis. Focusing on the monopole, we verified that, as expected, the RSDs cause an enhancement of the clustering signal at all scales, with respect to real space, quantifiable by a factor of $1.1-1.2$, due to the high mass and thus high bias of the haloes. Conversely, the effect of photometric errors is dominant, leading to a complete deletion of the BAO feature and to a scale-dependent suppression of the clustering signal, which is particularly important at low scales. This leads to a change in the slope of the correlation function with respect to the unperturbed one. The impact of photometric errors is even more important on the higher-order multipoles of the correlation function. Indeed, the quadrupole changes sign, providing a positive contribution, while the hexadecapole, which in redshift space is always compatible with zero for $r>20$ Mpc $h^{-1}$, becomes dominant. From the full set of mocks we measured also the numerical covariance matrix:
\begin{equation}
    \label{eq_numerical_covariance}
    \hat{C}_{abij}=\frac{1}{N_\mathrm{mocks}-1}\sum_{m=1}^{N_\mathrm{mocks}}
    (\xi^m_{ia}-\overline{\xi}_{ia})(\xi^m_{jb}-\overline{\xi}_{jb}),
\end{equation}
where $a$ and $b$ refer to the spatial bins involved, while $i$ and $j$ mark the considered redshift bins. The square root of the diagonal elements gives us the error bars shown in Fig. \ref{fig_3D_correlation_pinocchio_02}. By normalising the covariance matrix with its diagonal elements, we can obtain the correlation matrix:
\begin{equation}
    R_{abij}=\frac{\hat{C}_{abij}}{\sqrt{\hat{C}_{aaij}\hat{C}_{bbij}}}, 
\end{equation}
which is shown in Fig. \ref{fig_covariance_3D_pinocchio} for real space, redshift space and photo-$z$ space measurements. The cross-correlation between different redshift bins is compatible with zero even in the presence of photometric errors, which cause an overlap between the redshift distributions of the haloes, thus it is not reported and we only investigate the auto-correlation within the same redshift bin. In particular, if we focus on the block diagonal submatrices, which represent the auto-correlation of the multipoles between different radial bins, we find non-negligible off-diagonal terms that are large even when the bins are widely separated. This is a known feature, reflecting the fact that the 3D and angular correlation functions, in configuration space, are more correlated than their power spectrum counterparts in Fourier and spherical harmonic space \citep{Crocce11, Chan22}. However, this correlation decreases with both increasing multipoles and redshift. Moreover, while we see no significant difference between real and redshift space, due to the limited impact of RSDs, we find a significant reduction in the correlation with larger photometric errors, which is also common to the $w(\theta)$ covariance matrices (see Sect. \ref{sect_wtheta_measure}).\\
\indent On the other hand, from the inspection of the off-block diagonal terms, we can assess the cross-correlation between different multipoles. As expected, in real space, it is consistent with zero for every considered redshift bin, since there is no contribution from quadrupole and hexadecapole. In redshift and photo-$z$ space, the correlation progressively reduces at large $z$ and it is generally higher between adiacent multipoles. In particular, in presence of photometric errors, we find that positive deviations of $\xi_0(r)$ and $\xi_2(r)$ from their mean value are associated to positive deviations in $\xi_2(r)$ and $\xi_4(r)$, if $r_0<r_2$ or $r_2<r_4$, respectively, while for $r_0>r_2$ and $r_2>r_4$ we find an anti-correlation. 
\begin{figure*}[htbp]
\centering
\includegraphics[width=1\textwidth]{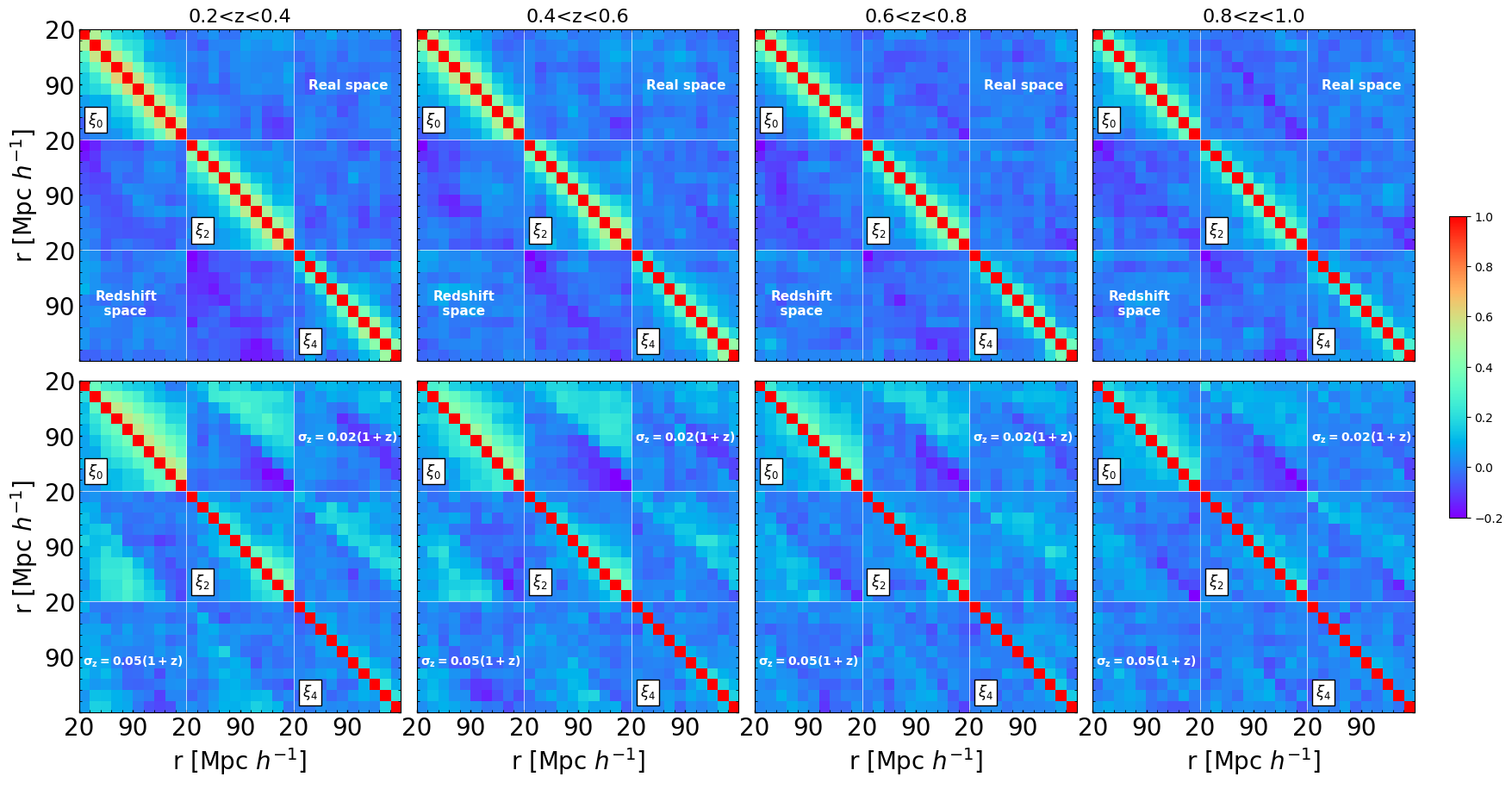}
\caption{Numerical correlation matrices of the 3D two-point correlation function first three even multipoles, derived from 1000 PLCs. The corresponding colour-bar is shown on the right. Top panels: the red diagonal separates the results for real space (upper triangles) and redshift space, with $\sigma_z=0$ (lower triangles). Bottom panels: the red diagonal separates the results in photo-$z$ space, where RSDs are included, for photometric errors given by $\sigma_z=0.02(1+z)$ (upper triangles) and $\sigma_z=0.05(1+z)$ (lower triangles). In each panel, horizontal and vertical white solid lines separate the correlation sub-matrices of the monopole, $\xi_0(r)$, the quadrupole, $\xi_2(r)$, and the hexadecapole, $\xi_4$, as well as their corresponding cross-correlations.}
\label{fig_covariance_3D_pinocchio}
\end{figure*}
\subsection{Modelling the 3D two-point correlation function}
Given the density contrast field $\delta(\boldsymbol{x})\equiv\delta\rho(\boldsymbol{x})/\rho_\mathrm{b}$, where $\rho_\mathrm{b}$ is the mean background density, it is possible to define the 3D two-point correlation function:
\begin{equation}
    \xi(\boldsymbol{x_1, x_2})=\langle \delta(\boldsymbol{x_1}) \delta(\boldsymbol{x_2}) \rangle, 
\end{equation}
as the expectation value, $\langle...\rangle$, of the product of the overdensities positioned at points $\boldsymbol{x_1}$ and $\boldsymbol{x_2}$. For the cosmological principle we can assume isotropic density fluctuations, so that $\xi(\boldsymbol{x_1, x_2})=\xi(r)$, i.e. it depends only on the distance between the sources, $r\equiv |\boldsymbol{x_2}-\boldsymbol{x_1}|$, and not on their positions. Dark matter haloes can be treated as biased tracers of the underlying matter distribution. Therefore, in the range of masses and linear scales that we are analysing, this translates into a linear relationship between the halo and the dark matter density contrast, $\delta_{\mathrm{h}}(\boldsymbol{x})$ and $\delta(\boldsymbol{x})$: 
\begin{equation}
\label{eq_biased_density}
  \delta_{\mathrm{h}}(\boldsymbol{x})= b_{\mathrm{h}}(z)\delta(\boldsymbol{x}),  
\end{equation}
through a scale-independent bias, $b_{\mathrm{h}}(z)$, which implies a halo-halo correlation function, $\xi_\mathrm{hh}(r)=b^2_{\mathrm{h}}(z)\xi(r)$. Since the power spectrum is the Fourier transform of the correlation function, the linear bias can be expressed in the same way, as the ratio between the power spectra: $P_\mathrm{hh}(k)=b^2_{\mathrm{h}}P(k)$.\\
\indent For $P(k)$ we adopt the linear model, but taking into account the nonlinear damping effects which produce a smearing of the two-point correlation function at the BAO peak, summarised in the parameter $\Sigma_\mathrm{NL}$, which is left free to vary. Therefore, following e.g. \citet{Veropalumbo16, Avila18} and \citet{Chan18}, we modelled $P(k)$ as follows:
\begin{equation}
\label{eq_Pk_IR_resummation}
    P(k)=[P_\mathrm{lin}(k)-P_\mathrm{nw}(k)]e^{-k^2\Sigma^2_{\mathrm{NL}}/2}+P_\mathrm{nw}(k),
\end{equation}
where $P_\mathrm{lin}(k,z)=P_\mathrm{lin}(k,0)D^2(z)$ is provided by CAMB \citep{Lewis00}, and is strictly valid only in linear theory \citep{Blake07}, $D(z)$ is the linear growth factor, normalised so that $D(0)=1$, while $P_\mathrm{nw}(k)$ is the power spectrum without the BAO peak, as obtained by the parametrisation presented in \citet{Eisenstein98}.
The real-space correlation function, $\xi(r)$, is then obtained by Fourier transforming the corresponding power spectrum.\\
\indent To compare $\xi_\mathrm{hh}(r)$ to the 3D correlation measured from the simulations, we shall consider the effective bias, i.e. the linear bias integrated above our mass threshold. In practice, $b_{\mathrm{eff}}$ is the halo bias weighted with the halo mass function:
\begin{equation}
    \label{eq_eff_bias}
    b_\mathrm{eff}(\geq M, z)=\frac{\int_M^\infty \mathrm{d}M\, \frac{dn}{dM}(M,z)b(M,z) }{\int_M^\infty \mathrm{d}M\,\frac{dn}{dM}(M,z)}.
\end{equation}
where $\frac{\mathrm{d}n(M,z)}{\mathrm{d}M}$ is computed with the parametrisation provided by \citet{Despali16}, which was also used for the calibration of the halo mass function
of the \textsc{Pinocchio} catalogues, as mentioned in Sect. \ref{sec_pinocchio_catalogue}. For $b(M,z)$ we adopt the theoretical halo-bias model presented in \citet{Tinker10}.
\subsubsection{Redshift-space and geometric distortions}
In real galaxy and cluster photometric surveys, the observed redshifts of the tracers, $z_\mathrm{obs}$, are linked to the cosmological redshifts, $z_\mathrm{c}$, as follows (neglecting all relativistic corrections and redshift measurement uncertainties):
\begin{equation}
\label{eq_obs_redshift}
    z_\mathrm{obs}=z_\mathrm{c}+\frac{v_\|}{c}(1+z_\mathrm{c}),
\end{equation}
where $z_\mathrm{c}$ is related to the cosmological Hubble flow, while the second term is due to the line-of-sight peculiar velocities, $v_\|$, which introduce distortions in the clustering signal if ignored. We neglect the so-called fingers-of-God (FoG) effect, generated by the nonlinear dynamics at small scales \citep{Jackson1972, Peacock01}, while we consider the large-scale squashing determined by the coherent infall towards collapsing structures, namely the Kaiser effect \citep{Kaiser84}. In the latter case, the linear regime velocity field can be directly determined from the density field and the anisotropic redshift space power spectrum can be parameterised as:
\begin{equation}
\label{eq_pk_kaiser}
    P_\mathrm{hh}(k, \mu)=b_\mathrm{eff}^2 \left(1+\frac{f}{b_\mathrm{eff}} \mu^2\right)^2 P(k),
\end{equation}
where $P(k)$ is computed with Eq. \eqref{eq_Pk_IR_resummation},  $f\equiv \mathrm{d}\ln D / \mathrm{d} \ln a$ is the linear growth rate and $\mu$ is the cosine of the angle between $\boldsymbol{k}$ and the line of sight. Here, the $f\mu^2$ term reproduces at all scales the coherent motions of large-scale structure, which acts as a bulk flows from underdense to overdense regions. RSDs in the Kaiser limit are proportional
to the parameter $\beta=f/b_\mathrm{eff}$, thus their effect is less important in more clustered, i.e. in large mass haloes. \\
\indent The anisotropic redshift-space 3D correlation function, $\xi_\mathrm{hh}(s,\mu)$, at the redshift-space comoving distance $s$, can be expressed in terms of multipoles $\xi_\ell(s)$ and Legendre polynomials $L_\ell(\mu)$ \citep{Hamilton92}:
\begin{equation}
\label{eq_xi_multipoles}
    \xi(s, \mu)=\xi_0(s)+\xi_2(s)L_2(\mu)+\xi_4(s)L_4(\mu)+\mathcal{O}(s^4).
\end{equation}
To model the multipoles, we can start from the expansion of the anisotropic power spectrum in Eq. \eqref{eq_pk_kaiser}: 
\begin{equation}
    \label{eq_pkl_multipoles}
    P_{\ell}(k)=\frac{2 \ell+1}{2\alpha_{\perp}^2 \alpha_{\parallel}} \int^{+1}_{-1} \mathrm{d} \mu P\left(k', \mu'\right) L_{\ell}\left(\mu\right), 
\end{equation}
where $\mu$ is the line-of-sight cosine, while $k'$ and $\mu'$ are rescaled according to the prescription of \citet{Beutler14}:
\begin{equation}
    k^{\prime}=\frac{k}{\alpha_{\perp}}\left[1+\mu^2\left(\frac{\alpha_{\perp}^2}{\alpha_{\|}^2}-1\right)\right]^{1 / 2},
\end{equation}
and: 
\begin{equation}
    \mu^{\prime}=\mu \frac{\alpha_{\perp}}{\alpha_{\|}}\left[1+\mu^2\left(\frac{\alpha_{\perp}^2}{\alpha_{\|}^2}-1\right)\right]^{-1 / 2}.
\end{equation}
The geometric
distortions are corrected through: 
\begin{equation}
\begin{aligned}
\alpha_{\parallel} &= \frac{H^{\mathrm{fid}}(z)}{H(z)} \frac{r_s^{\mathrm{fid}}\left(z_d\right)}{r_s(z_d)}, \\
\alpha_{\perp} &= \frac{D_A(z)}{D_A^{\mathrm{fid}}(z)} \frac{r_s^{\mathrm{fid}}\left(z_d\right)}{r_s\left(z_d\right)},
\end{aligned} 
\end{equation}
where $H^\mathrm{fid}(z)$ and $D_A^\mathrm{fid}(z)$ represent the fiducial Hubble parameter and angular diameter distance, respectively, while $r_s(z_d)$ is the sound horizon at the drag redshift, $z_d$, and $r_s^\mathrm{fid}(z_d)$ is its fiducial value. Then, the multipoles $\xi_\ell$ are expressed as follows: 
\begin{equation}
    \label{eq_xil_multipoles}
   \xi_{\ell}(s)=\frac{i^{\ell}}{2 \pi^2} \int \mathrm{d} k k^2 P_{\ell}(k) j_{\ell}(k s),
\end{equation}
where $i$ is the imaginary unit and $j_\ell(x)$ is the $\ell$-th order spherical Bessel function. The monopole can be simply written as a function of the real-space correlation function, $\xi(r)$:
\begin{equation}
    \xi_{\mathrm{hh,}0}(s)=\left[b^2_{\mathrm{eff}}+\frac{2}{3}b_{\mathrm{eff}}f+\frac{1}{5}f^2 \right]\xi(\alpha r), 
\end{equation}
where $\alpha$ is the parameter that corrects for the geometric distortions:
\begin{equation}
    \alpha = \frac{D_V(z)}{D_V^\mathrm{fid}(z)} \frac{r_s^\mathrm{fid}(z_d)}{r_s(z_d)},
\end{equation}
where $D_V$ is the average distance at redshift $z$, and $D_V^\mathrm{fid}$ is the same quantity, but estimated at the fiducial cosmology. Their ratio rescales the distances at which the correlation function model is evaluated.
\subsubsection{Photometric errors}
\label{sect_photometric_errors}
Redshift errors introduce a $\sigma_z$ term in Eq. \eqref{eq_obs_redshift}, perturbing the distance measurements along the line of sight:
\begin{equation}
    z_\mathrm{obs}=z_\mathrm{c}+\frac{v_\|}{c}(1+z_\mathrm{c})\pm \sigma_z.
\end{equation}
Following the approach presented in \citet{Blake05, Asorey12, Pezzotta17}, we can modify Eq. \eqref{eq_pk_kaiser} as: 
\begin{equation}
\label{eq_pk_photoz_errors}
    P_\mathrm{hh}(k, \mu)=P(k)\, b_\mathrm{eff}^2 \left(1+\frac{f}{b_\mathrm{eff}} \mu^2\right)^2 \mathrm{exp}(-k^2\mu^2\sigma^2).
\end{equation}
The exponential factor in Eq. \eqref{eq_pk_photoz_errors} affects the clustering signal in a way similar to that of random peculiar motions \citep{Marulli12}. Indeed, it translates in a Gaussian damping term,
analogous to the FoG effect \citep{GarciaFarieta20},
which causes a scale-dependent suppression of the signal for $k>1/\sigma$, where $\sigma$ quantifies the impact of photometric random perturbations on cosmological
redshifts. In particular: 
\begin{equation}
    \sigma=\frac{c\sigma_z}{H(z)},
\end{equation}
where $c$ is the speed of light and $H(z)$ is the Hubble parameter. This is equivalent to assuming that the redshift errors follow a Gaussian distribution with zero mean and with dispersion given by $\sigma_z$ \citep{Hutsi10}.\\
\indent The monopole of the 3D correlation function can be computed after integrating over $\mu$, following the redshift-space model presented in \citet{Sereno15}, and used by \citet{Lesci22}:
\begin{equation}
\label{eq_xi_damped_sereno}
    \xi_{\mathrm{hh},0}(s) = b_\mathrm{eff}^2\xi'(s) + b_\mathrm{eff}\xi''(s) + \xi''' (s), 
\end{equation}
where different terms of the correlation function are obtained from the Fourier anti-transforms of the corresponding power spectra. In particular: 
\begin{equation}
    P'(k)=P(k)\frac{\sqrt{\pi}}{2k\sigma}\mathrm{erf}(k\sigma),
\end{equation}
\begin{equation}
    P''(k)=\frac{f}{(k\sigma)^3}P(k)\left[ \frac{\sqrt{\pi}}{2}\mathrm{erf}(k\sigma)-k\sigma \mathrm{exp}(-k^2\sigma^2)\right],
\end{equation}
and
\begin{multline}
    P'''(k) = \frac{f^2}{(k\sigma)^5} P(k) \left\{ \frac{3\sqrt{\pi}}{8}\mathrm{erf}(k\sigma) \right.+ \\
    - \left. \frac{k\sigma}{4} [2(k\sigma)^2+3] \mathrm{exp}(-k^2\sigma^2) \right\},
\end{multline}
respectively. We notice that in real space $ P''(k)$ and $ P'''(k)$ are both zero. Finally, the multipoles are simply computed by using the power spectrum of Eq. \eqref{eq_pk_photoz_errors} into the Eqs. \eqref{eq_pkl_multipoles} and \eqref{eq_xil_multipoles}.
\section{Angular two-point correlation function}
\label{sect_angular_correlation_function}
\subsection{Characterisation of the measurement and covariance matrix}
\label{sect_wtheta_measure}
\begin{figure}[htpb]
\centering
\includegraphics[width=\hsize]{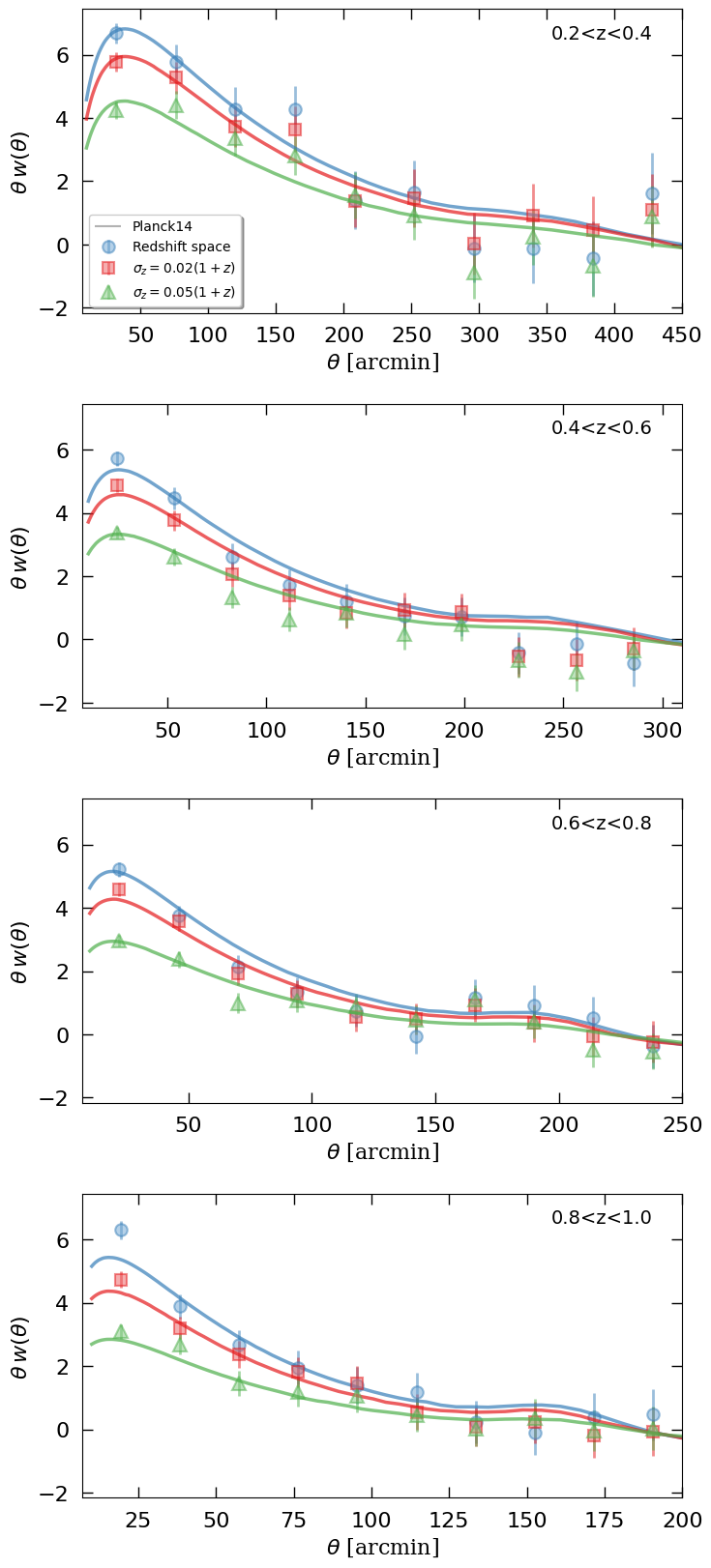}
\caption{The angular two-point correlation function, measured from one \textsc{Pinocchio} PLC. Errors are computed as the square root of the diagonal elements of the covariance matrix. We show results for redshift space, with $\sigma_z=0$ (blue circles), and photo-$z$ space with RSDs, for photometric errors equal to $\sigma_z=0.02(1+z)$ (red squares), and $\sigma_z=0.05(1+z)$ (green triangles). The solid lines show the theoretical two-point correlation function, computed from the power spectrum estimated with CAMB, assuming a linear model with $\Sigma_\mathrm{NL}=0$, i.e. without nonlinear damping effects at the BAO scale. They include RSDs and their colours corresponds to the photometric damping indicated by the symbols.}
\label{fig_wtheta_correlation_pinocchio_02}
\end{figure}

\begin{figure*}[htbp]
\centering
\includegraphics[width=1\textwidth]{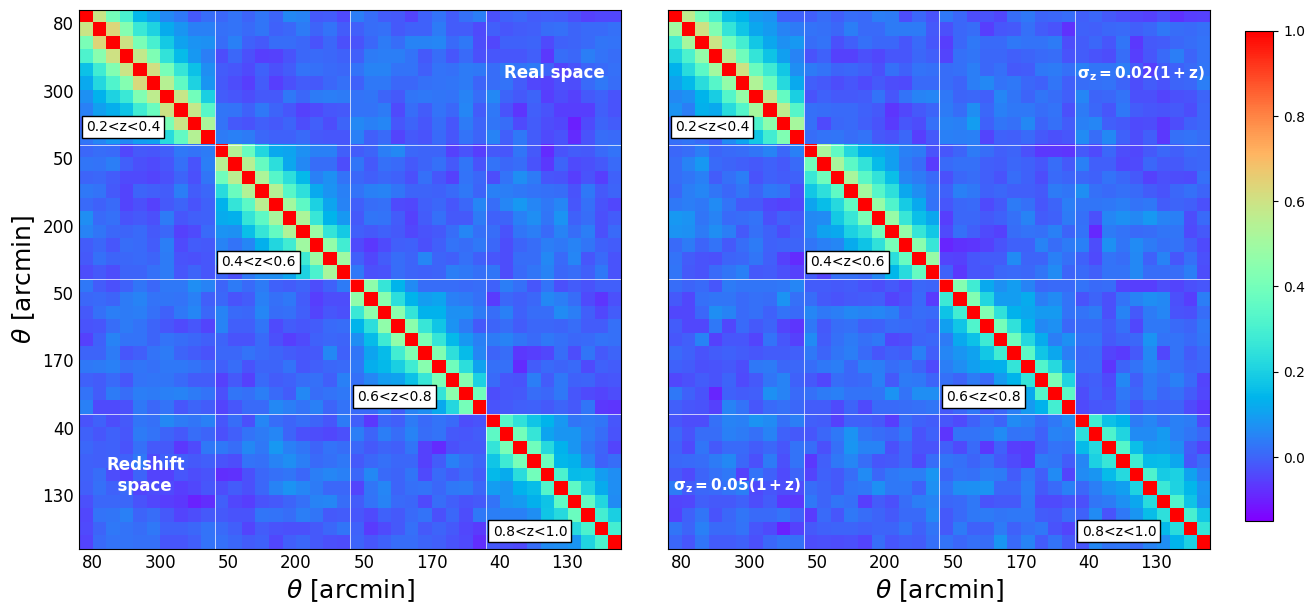}
\caption{Numerical correlation matrices of the angular two-point correlation function, derived from 1000 PLCs. The corresponding colour-bar is shown on the right. Left: results for real space (upper triangle) and redshift space, with $\sigma_z=0$ (lower triangle). Right: results in photo-$z$ space, where RSDs are included, for photometric errors given by $\sigma_z=0.02(1+z)$ (upper triangle) and $\sigma_z=0.05(1+z)$ (lower triangle). In each panel, horizontal and vertical white solid lines separate the cross-correlation sub-matrices measured between different redshift bins.}
\label{fig_covariance_wtheta_pinocchio}
\end{figure*}

The angular correlation function measures the average correlation between any two objects placed in the solid angle elements $\delta \Omega_1$ and $\delta \Omega_2$, separated by an angle $\theta$. It can be defined in a completely analogous way with respect to its full 3D counterpart:   
\begin{equation}
    \delta P_{12}(\theta)= n_{\Omega}^2[1+w(\theta)]\delta \Omega_1 \delta \Omega_2,
\end{equation}
where $n_\Omega$ is the mean number of objects per unit solid angle. We measure $w(\theta)$ in 10 linearly-spaced bins with the LS estimator \citep{Landy93}:
\begin{equation}
    w_{\mathrm{LS}}(\theta)=\frac{DD(\theta)+RR(\theta)-2DR(\theta)}{RR(\theta)},
\end{equation}
where $DD(\theta)$, $RR(\theta)$, and $DR(\theta)$ are the normalised number of data-data, random-random, and data-random pairs in the angular bin $\theta \pm \Delta \theta/2$, respectively. The advantage of the angular correlation measurement is that it requires only the random catalogue in RA and Dec, without any information about the redshift distribution. This allows us to greatly simplify the measurement process.\\
\indent We choose angular ranges that vary as a function of redshift. In particular, we explore scales above the maximum angular size obtained from the virial radii of the halo sample, while for $z>0.4$ we adopt a more conservative limit of $20$ arcmin, in order to exclude comoving separations below $10$ Mpc $h^{-1}$, where there is a significant deviation of the clustering signal from the linear model prediction, both in $\xi(r)$ and in $w(\theta)$. In Fig. \ref{fig_wtheta_correlation_pinocchio_02} we show our results for each redshift bin. For graphical requirements, we plot $\theta\,w(\theta)$ and we set independent x-axes. In fact, the same angle covers greater physical distances as the redshift increases. This results in a shift of $w(\theta)$ towards smaller scales. As for the 3D case, the RSDs have only a slight impact. Again, we find the dominant effect in the photo-$z$ space, where photometric errors cause a drop in the angular clustering. \\
\indent It is noteworthy that, in contrast to the 3D case, the BAO feature is preserved, suffering only a minor degradation when passing to $\sigma_{z,0}=0.02-0.05$. This is because redshift uncertainties affect only the radial distances and not the transverse angular ones. Thus, their effect is greater on the reconstruction of the 3D correlation function or power spectrum, which depend on $r$ or $k$, rather than on their angular counterparts. \\
\indent In Fig. \ref{fig_covariance_wtheta_pinocchio} we show the numerical covariance matrices, derived from the full set of 1000 PLCs. Their features are similar to the 3D case: no significant level of correlation between different redshift bins, even after the introduction of photo-$z$ errors; positive auto-correlation between different radial bins, which decreases progressively with redshift and with larger photometric errors. 
\subsection{Modelling the angular two-point correlation function}
\label{Sec_wtheta_model_pinocchio}
The study of angular clustering considers the projection of the spatial halo density field onto the celestial sphere, in a given direction of the sky $\hat{\boldsymbol{n}}$:
\begin{equation} 
\label{eq_delta_projected}\delta_{\mathrm{h}}^i(\hat{\boldsymbol{n}})=\int \mathrm{d}z \, \phi^i(z) \delta_{\mathrm{h}}^i (\boldsymbol{x}),
\end{equation}
where $\phi^i(z)$ is the normalised selection function  in the $i$-th redshift bin. Given the projected density field, the angular correlation function can be defined as \citep{Peebles73}:
\begin{equation}
    w_\mathrm{hh}(\theta) = \langle \delta_\mathrm{h}(\hat{\boldsymbol{n}}) \delta_\mathrm{h}^*(\hat{\boldsymbol{n}}+\hat{\boldsymbol{\theta}}) \rangle,
\end{equation}
and can be computed as the projection of the 3D spatial correlation function, $\xi(s)$:
\begin{equation}
\label{eq_wtheta_model}
    w_\mathrm{hh}^{ij}(\theta)=\int \int \mathrm{d}z_1 \mathrm{d}z_2 \phi^i(z_1) \phi^j(z_2) \xi_\mathrm{hh}(s,\mu),
\end{equation}
where $s=\sqrt{r^2(z_1)+r^2(z_2)-2r(z_1)r(z_2)\cos\theta}$, $\mu$ is the same of Eq. \eqref{eq_pk_kaiser} and takes the form $\mu=\frac{r(z_2)-r(z_1)}{s}\cos (\theta/2)$, and $r(z)$ is the comoving distance at redshift $z$. The case $i=j$ refers to the auto correlation, while $i\neq j$ to the cross correlation between different redshift bins. RSDs can be naturally incorporated, by including a multiplicative factor at the level of the power spectrum. \\
\indent The anisotropic redshift-space 3D correlation function, $\xi_\mathrm{hh}(s,\mu)$ can be expressed with an infinite series of even multipoles, $\xi_\ell(s)$, and Legendre polynomials, $L_\ell(\mu)$, such that $\xi(s,\mu)=\sum_\ell \xi_\ell(s) L_\ell(\mu)$, 
as in Eq. \eqref{eq_xi_multipoles}. Most of the information is contained in the monopole \citep{GarciaFarieta20}, but following \citet{Salazar14}, to avoid systematics, we include also the correction given by the quadrupole. We do not account for successive terms, because we have verified that the hexadecapole contribution is negligible, for the adopted $P(k)$ model. 
\subsection{Redshift selection function model}
\label{Sect_Selection function}
\begin{figure}[htpb]
\centering
\includegraphics[width=\hsize]{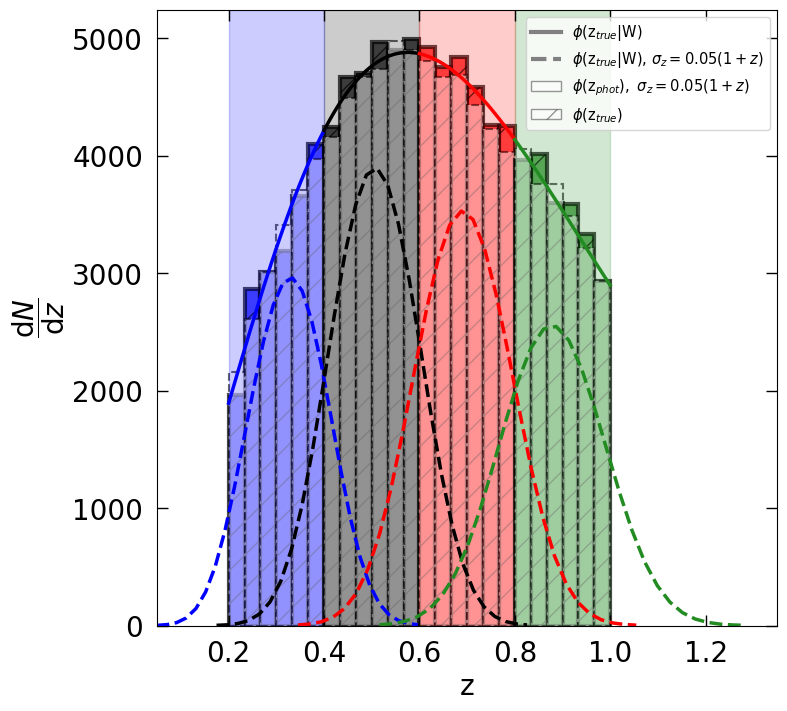}
\caption{The redshift distribution of dark matter haloes, $\mathrm{d}N /\mathrm{d}z$, from a single realisation of \textsc{Pinocchio} PLC, with an angular radius of approximately $60$ deg, in the redshift range $0.2<z<1.0$. The coloured hatched histogram represented with a thick solid line shows the real-space redshift distribution, while the dashed histogram shows the photometric distribution, after the introduction of a Gaussian perturbation with $\sigma_z=0.05(1+z)$. The light coloured shaded regions sign the limits of the photometric redshift bins of our tomographic analysis, with $\Delta z=0.2$.
The solid line gives the true $\mathrm{d}N/\mathrm{d}z$ in the window function $W$, derived by integrating the halo mass function of \citet{Despali16}, assuming the cosmological parameters of the simulation. The dashed lines are derived through Eq. \eqref{eq_dN_dzi_convolution} and show the expected halo distribution in the $i$-th redshift bin, selected by $W$.
}
\label{fig_redshift_distribution_pinocchio}
\end{figure}

The radial selection function, $\phi(z)$, encodes the probability of including a halo in a given redshift bin. In the absence of redshift errors, $\phi(z)$ reduces to the true underlying redshift distribution:
\begin{equation}
    \phi(z)=\frac{1}{N}\frac{\mathrm{d}N}{\mathrm{d}z}, 
\end{equation}
where:
\begin{equation}
\label{eq_theoretical_true_dN_dz}
    \frac{\mathrm{d}N}{\mathrm{d}z}=\Omega_{\mathrm{sky}} \frac{\mathrm{d}V}{\mathrm{d}z \mathrm{d} \Omega} \int_{0}^{\infty} \frac{\mathrm{d}n(M, z)}{\mathrm{d}M} \mathrm{d}M,
\end{equation}
which is related to the survey area, $\Omega_{\mathrm{sky}}$, in our case $10\,313$ deg$^2$. Eq. \eqref{eq_theoretical_true_dN_dz} has a cosmological dependence given by the derivative of the comoving volume, $\frac{\mathrm{d}V}{\mathrm{d}z \mathrm{d} \Omega}$, and by the halo mass function, $\mathrm{d}n(M,z)/\mathrm{d}M$, for which we use the functional form provided by \citet{Despali16}. \\
\indent The selection function in a given redshift bin is determined by the window function, $W(z)$:
\begin{equation}
    \phi^i(z)=\phi(z|W)=\phi(z)W(z),
\end{equation}
which in our redshift-space mock can simply be considered as a top-hat function in the chosen redshift shell. Complications arise when we consider photometric redshifts, which have non-negligible errors that can significantly alter the clustering signal. Indeed, when dealing with angular clustering, redshift uncertainties enter into the redshift selection function as a modification of the halo redshift distribution, rather than as a damping term at the level of the power spectrum, as we have seen in Sect. \ref{sect_photometric_errors} \citep{Asorey12}. So we need to consider the conditional probability $P(z|z_{\mathrm{phot}})$ of having a halo at the true redshift, $z$, given the photometric redshift, $z_{\mathrm{phot}}$. The selection of an object in a photometric redshift bin is determined by a top-hat window function in $z_\mathrm{phot}$, reflecting the adopted binning strategy \citep{Asorey12}:
\begin{equation}
    W(z_{\mathrm{phot}})=  \begin{cases}
    0 \qquad  z_{\mathrm{phot}}\leq z^i_{\mathrm{min}} \ \mathrm{or} \ z_{\mathrm{phot}}>z^i_{\mathrm{max}} \\
    1 \qquad  z^i_{\mathrm{min}}< z_{\mathrm{phot}} \leq z^i_{\mathrm{max}} \\
\end{cases}
,
\end{equation}
where $z_{\mathrm{min}}$ and $z_{\mathrm{max}}$ represent the limits of our photometric redshift interval, respectively.\\ 
\indent Thus, the normalised redshift distribution in the $i$-th photometric bin is obtained with the following convolution \citep{Budavari03, Crocce11, Hutsi14}:
\begin{equation}
\label{eq_selection_function}
    \phi^i(z)=\phi(z|W)=\phi(z)\int_0^\infty \mathrm{d}z_{\mathrm{phot}}W(z_{\mathrm{phot}})P(z_{\mathrm{phot}}|z),
\end{equation}
where $P(z_{\mathrm{phot}}|z)$ is given by Eq. \eqref{eq_gaussian_perturbation_photoz}. This quantity can assume a complicated form, depending on several parameters, including the fraction of catastrophic outliers, $f_\mathrm{out}$, namely systems with severely misestimated photometric redshifts \citep[e.g.][]{Hutsi14, Blanchard20}. For example, in optically selected clusters, if the catastrophic outlier fraction of the member galaxies is large, the cluster properties could be altered, as an increased number of false detections or a large uncertainty in the redshift estimate \citep{Adam19}. However, in our case, for construction, a simple Gaussian distribution is assumed, whose mean is $z$, while the standard deviation is equal to $\sigma_z=\sigma_{z,0}(1+z)$, with $\sigma_{z,0} = 0-0.02-0.05$, depending on the photometric error of our mocks. This choice was also adopted in \citet{Romanello24} and is motivated by the fact that we do not expect a significant shift in the estimated cluster redshift of real catalogues, assuming the outlier rate of recent photometric surveys \citep[e.g.][]{Hildebrandt17}. Finally, the redshift distribution in the $i$-th redshift bin is expressed by: 
\begin{equation}
\label{eq_dN_dzi_convolution}
\begin{split}
   \frac{dN}{dz^i}=\Omega_{\mathrm{sky}} \frac{\mathrm{d}V}{\mathrm{d}z \mathrm{d} \Omega} \int_{0}^{\infty} \mathrm{d}M \frac{\mathrm{d}n(M, z)}{\mathrm{d}M}
    \int_{\Delta z_i} \mathrm{d} z_{\mathrm{phot}}P(z_{\mathrm{phot}}|z). 
\end{split}
\end{equation}
In Fig. \ref{fig_redshift_distribution_pinocchio} we show the redshift distribution of haloes, in $0.2<z<1.0$. For the sake of simplicity, we report only two extreme cases. The coloured, hatched histogram, represented by a solid thick line, shows the real-space redshift distribution, while the dashed histogram shows the photo-$z$ space redshift distribution, in which the redshifts are perturbed with $\sigma_{z,0}=0.05$. Due to this perturbation, some haloes leave their redshift bin, while others enter it. The net effect would be zero in the case of a uniform initial distribution in $z$. However, given the initial shape of the $\mathrm{d}N/\mathrm{d}z$ (solid line in Fig. \ref{fig_redshift_distribution_pinocchio}), that can be modelled by integrating the \citet{Despali16} mass function with Eq. \eqref{eq_theoretical_true_dN_dz}, between the minimum and maximum masses of the mock catalogue, the introduction of photometric errors determines a flattening of the overall distribution, leading to a migration of haloes from the most populated to the less populated bins. The result is shown with the partially overlapping dashed curves in Fig. \ref{fig_redshift_distribution_pinocchio}.
\section{Angular power spectrum}
\label{Sect_angular_power_spectrum}
\subsection{\textsc{Healpix} pixelated map}
\label{Sect_angular_power_spectrum_map_pinocchio}
Given the angular position and redshift of the haloes, we generate tomographic density maps, using the \textsc{Healpix} pixelisation scheme \citep{Gorski05}, with a resolution of $N_\mathrm{side} = 512$. The smoothing of the density field within equal area pixels results in a loss of information for $\ell\gtrsim 1500$, namely the corresponding pixel size. However, this value is well above the upper limits in which the measurements are made, i.e. $\ell = 150-200$, depending on the redshift bin. In each pixel, the density contrast, $\delta_{\mathrm{h}, p}$, is measured as:
\begin{equation}
    \delta_{\mathrm{h}, p}=\frac{n_{\mathrm{h},p}}{\Bar{n}_{\mathrm{h}}}-1,
\end{equation} 
where $n_{\mathrm{h},p}$ is the halo count in the $p$-th pixel, while $\Bar{n}_{\mathrm{h}}$ is the average halo count computed within the circle of $60$ degrees angular radius that delimits our PLCs.
\subsection{Angular power spectrum estimator}
\begin{figure}[htpb]
\centering
\includegraphics[width=0.95\hsize]{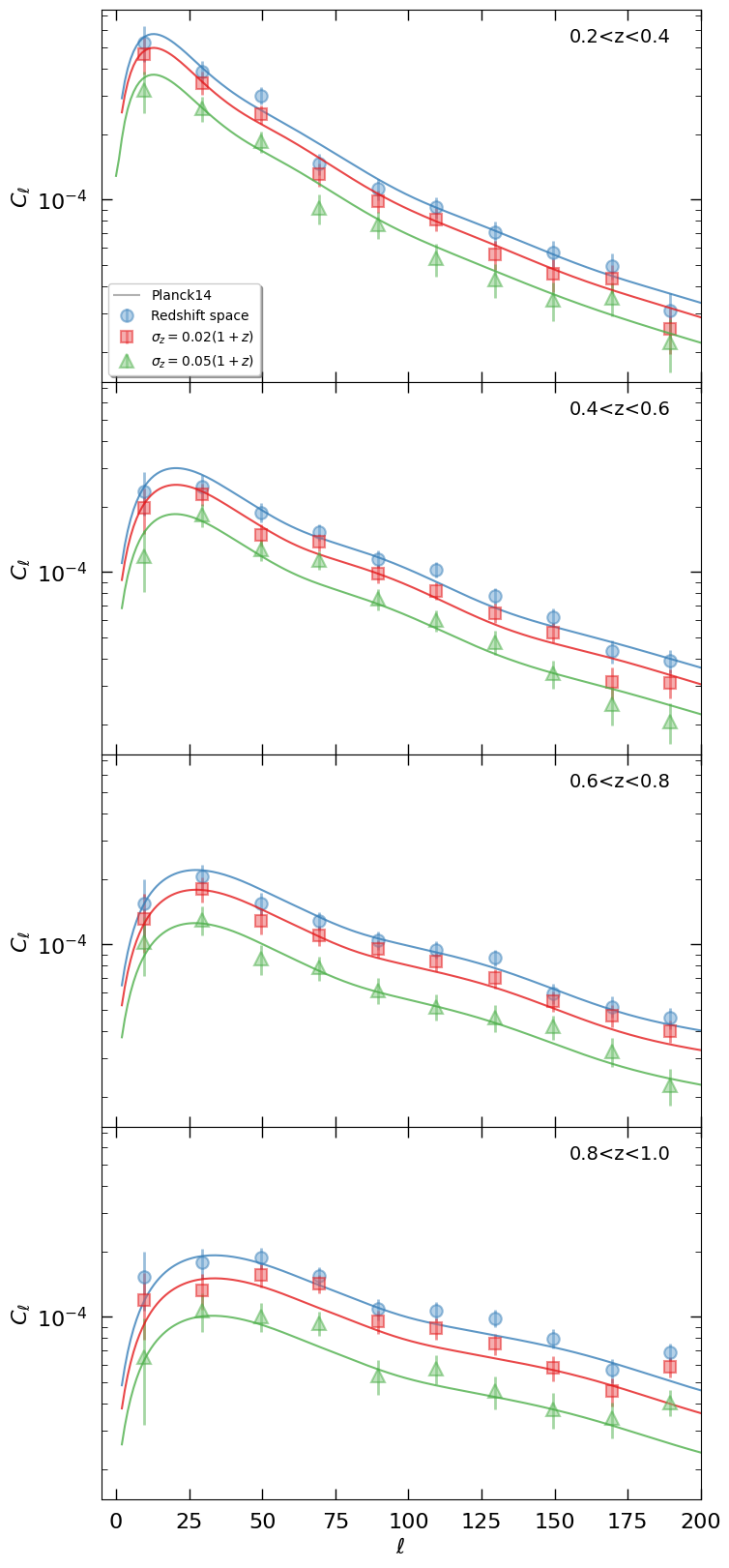}
\caption{The angular power spectrum, measured from one \textsc{Pinocchio} PLC. Errors are computed as the square root of the diagonal elements of the covariance matrix. We show results for redshift space, with $\sigma_z=0$ (blue circles), and photo-$z$ space with RSDs, for photometric errors equal to $\sigma_z=0.02(1+z)$ (red squares), and $\sigma_z=0.05(1+z)$ (green triangles). The solid lines show the theoretical angular power spectrum, computed from the power spectrum estimated with CAMB, assuming a linear model with $\Sigma_\mathrm{NL}=0$, i.e. without nonlinear damping effects at the BAO scale. They include RSDs and their colours corresponds to the photometric damping indicated by the symbols.}
\label{fig_Cl_pinocchio_02}
\end{figure}

\begin{figure*}[htbp]
\centering
\includegraphics[width=1\textwidth]{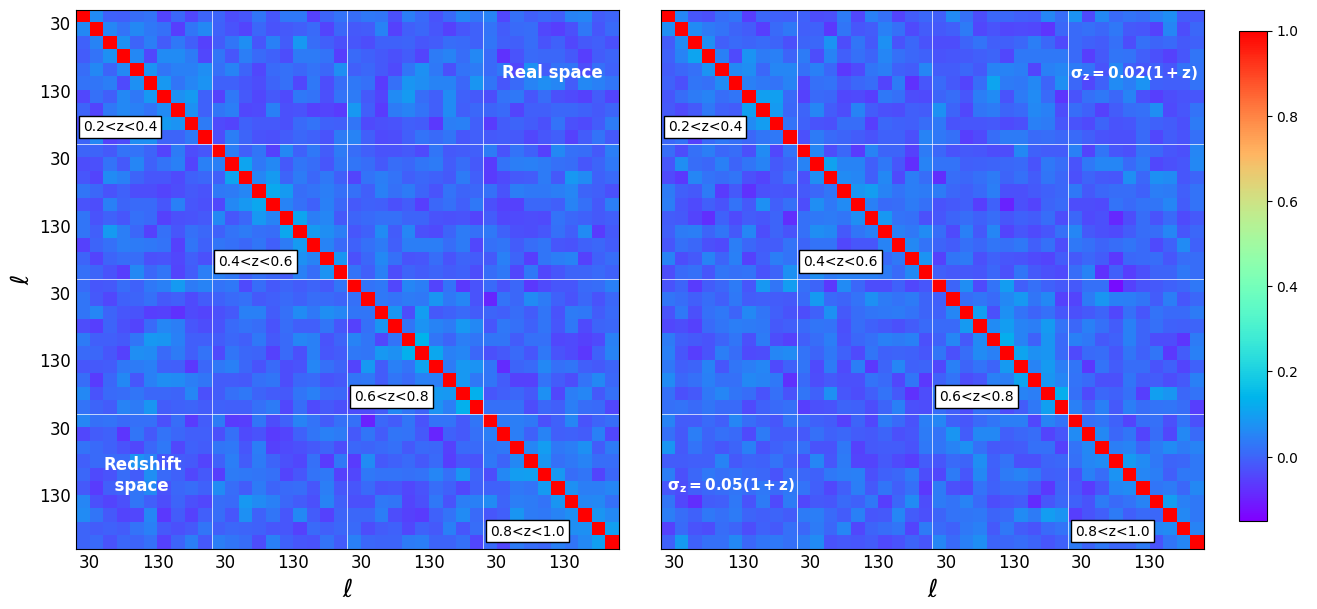}
\caption{Numerical correlation matrices of the angular power spectrum, derived from 1000 PLCs. The corresponding colour-bar is shown on the right. Left: results for real space (upper triangle) and redshift space, with $\sigma_z=0$ (lower triangle). Right: results in photo-$z$ space, where RSDs are included, for photometric errors given by $\sigma_z=0.02(1+z)$ (upper triangle) and $\sigma_z=0.05(1+z)$ (lower triangle). In each panel, horizontal and vertical white solid lines separate the cross-correlation sub-matrices measured between different redshift bins. 
}
\label{fig_Covariance_cl_pinocchio}
\end{figure*}

The angular power spectrum can be obtained from the projected density field expressed in Eq. \eqref{eq_delta_projected}. The density contrast, being defined on a 2D sphere, can be expanded into a series of spherical harmonics, $Y_{\ell m}$, computed at the direction on the sky $\hat{\boldsymbol{n}}\equiv (\theta,\varphi)$:
\begin{equation}      
\label{eq_spherical_harmonic_decomposition}
\delta^i_{\mathrm{h}}(\boldsymbol{\hat{n}})=\sum_{\ell=0}^{\infty}\sum_{m=-\ell}^{m=+\ell} a^i_{\ell m} Y_{\ell m}(\boldsymbol{\hat{n}}), 
\end{equation}
where $a_{\ell m}$ are the harmonic coefficients. The orthonormality of the $Y_{\ell m}$ implies that the last equation can be reversed, and approximated for a pixellised map:
\begin{equation}
    a^i_{\ell m}= \int \mathrm{d}\boldsymbol{\hat{n}}\, \delta^i_{\mathrm{h}}(\boldsymbol{\hat{n}}) Y^{*}_{\ell m}(\boldsymbol{\hat{n}}) \simeq \sum^{N_\mathrm{pix}}_p \delta_{\mathrm{h}, p}^iY^{*}_{\ell m}(\theta_p, \varphi_p)\Delta \Omega_p.
\end{equation}
The angular power spectrum can then be easily evaluated with an estimator already introduced by \citet{Hauser73, Peebles73} and widely used in the literature \citep[e.g.][]{Balaguera18}: 
\begin{equation}
    \label{eq_estimator_Cl}
    K^{ij}_\ell=\frac{1}{w^2_\ell}\left[\frac{1}{f_{\mathrm{sky}}(2\ell+1)}\sum_{m=-\ell}^{+\ell} |a^i_{\ell m}a^{*j}_{\ell m}|- \frac{\Omega_\mathrm{sky}}{N_{\mathrm{h}}}\delta^{ij}_K\right],
\end{equation}
where $w_\ell$ is the \textsc{Healpix} pixel window function, which corrects for the loss of power due to the pixelisation, $f_{\mathrm{sky}}$ is the sky fraction covered by our PLC, $N_{\mathrm{h}}$ is the number of haloes in the considered redshift bin and $\delta^{ij}_K$ is the Kronecker delta that cancels the shot noise term in the cross-correlation case.\\
\indent The results are shown in Fig. \ref{fig_Cl_pinocchio_02}. Coloured symbols refer to the measurements in redshift and photo-$z$ space, while error bars are estimated as the square root of the diagonal of the covariance matrix. RSDs have an impact that increases with redshift, and affect the measured angular power spectrum at large angular scales, corresponding to $\ell<10-40$, depending on the redshift bin. However, since we select haloes with large virial masses, their effect is generally negligible in the angular range of our analysis. On the other hand, photometric errors cause a general suppression of the clustering signal and need to be modelled in the same way as for the angular correlation function. We have averaged the $C_\ell$s in bands of $\Delta\ell=20$, which allows us to make the covariance matrix more diagonal, since it is blurred due to the partial sky coverage, and to reduce its size \citep{Blake07, Balaguera18, Loureiro19}. In Fig. \ref{fig_Covariance_cl_pinocchio} we show the correlation matrices, estimated by substituting $C_\ell$s in Eq. \eqref{eq_numerical_covariance}. As expected, assuming that the $a_{\ell m}$s follow a Gaussian distribution, different multipoles of the power spectrum are independent, and the correlation in spherical harmonics space is lower than in configuration space (see Appendices \ref{sec_covariance_cl} and \ref{sec_covariance_wtheta}).

\subsection{Shot noise correction}
\label{sect_shot_noise_pinocchio}
\begin{figure*}[htbp]
\centering
\includegraphics[width=1\textwidth]{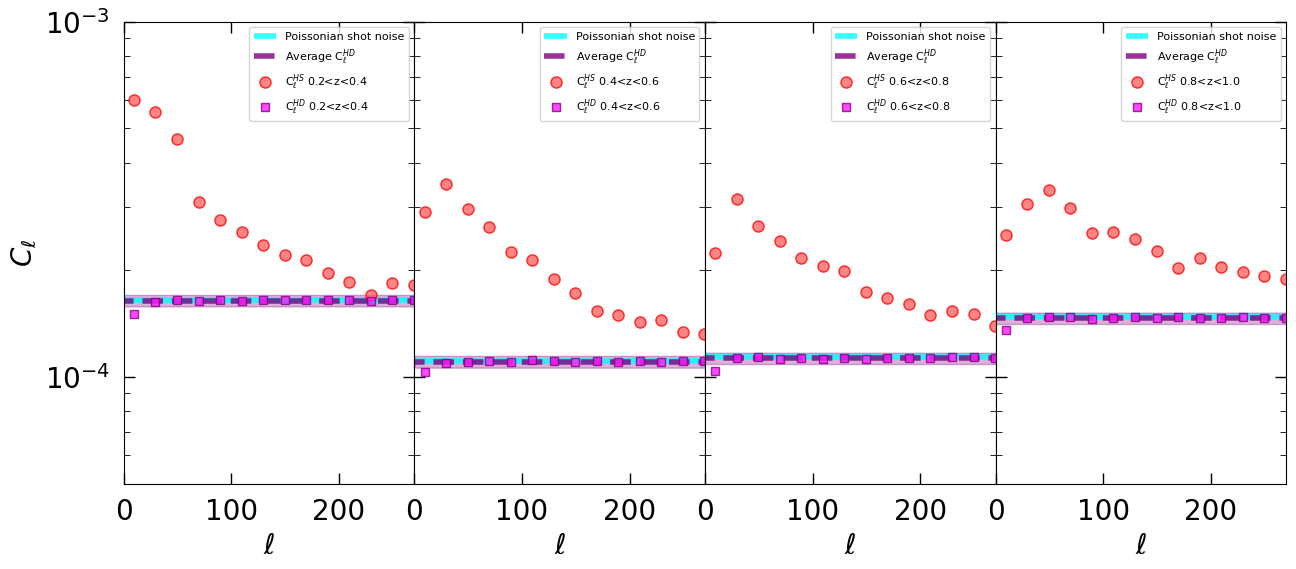}
\caption{Shot noise estimation over 100 realisations, using one realisation of the \textsc{Pinocchio} PLCs, in four redshift bins between $0.2<z<1.0$ (panels from left to right). The red circles represent the angular power spectrum of the half-sum (HS) maps, which include the contribution of both signal and noise. The purple squares instead show the angular power spectrum of the half-difference (HD) maps, which provide a direct estimate of the shot noise, with their average (purple dashed line) and standard deviation (purple shaded band) in agreement with the theoretical Poissonian value (cyan solid line).}
\label{fig_Check_shot_noise_Pinocchio_n512}
\end{figure*}

\indent The second term in the right-hand side of Eq. \eqref{eq_estimator_Cl} quantifies the part of the measured signal related to the discreteness distribution of tracers, which does not contain clustering information. This is called shot noise and must be removed in order to correctly estimate the angular power spectrum. We have seen that the shot noise can be modelled as the ratio between the survey area $\Omega_\mathrm{sky}$ and the number of haloes $N_\mathrm{h}$ in a given redshift bin, while it is equal to zero in the cross-correlation case.\\ 
\indent Depending on the number of objects, the shot noise is sensitive to the adopted binning strategy. It plays a minor role in large galaxy catalogues, while it generally becomes important when working with smaller catalogues of clusters. In our case, due to the $M>10^{14} M_\odot \, h^{-1}$ mass cut, we have approximately $10^4$ haloes in each redshift bin, between $0.2<z<1.0$. This fact enhances the relative importance of the shot noise, which becomes completely dominant over the signal even at relatively large angular scales. For this reason, we set the upper limit of our measurements at the scale where it is about twice as large as the signal, which corresponds to $\ell=150-200$, depending on the redshift bin. These values are equivalent to the lower limits $\theta=1.2$ deg and $0.9$ deg, respectively. Thus, the final angular range of the $C_\ell$ analysis is smaller than that considered for $w(\theta)$.\\
\indent The Poissonian approximation assumes point-like objects and so may not hold for finite-size tracers, such as haloes. In fact, we expect mass-dependent deviations from $\Omega_\mathrm{sky}/N_\mathrm{h}$, due to the halo exclusion, namely the fact that it is not possible for the distance between two haloes to be smaller than the sum of their physical sizes, and nonlinear effects \citep{Giocoli10, Baldauf13, Paech17}. In \citet{Romanello24} we verified that this approximation holds for galaxy clusters in the angular range $\ell \in [10-175]$, which roughly corresponds to the angular scales we are focusing on. In particular, we use a method already introduced in \citet{Ando18}, \citet{Makiya18} and \citet{Ibitoye22}, for the analysis of the galaxy-galaxy angular power spectrum. In practice, we randomly split our PLC into two sub-catalogues each containing, by construction, roughly the same number of dark matter haloes. Then we build the corresponding density maps, $\delta_{1,\mathrm{h}}$ and $\delta_{2,\mathrm{h}}$, which are slightly different in each realisation, being based on a random extraction process. By combining them, we can build the half-sum, $\mathrm{HS}=\frac{1}{2}(\delta_{1,\mathrm{h}}+\delta_{2,\mathrm{h}})$, and the half-difference density fluctuation maps, $\mathrm{HD}=\frac{1}{2}(\delta_{1,\mathrm{h}}-\delta_{2,\mathrm{h}})$. The half-sum map contains both signal and noise, while only the shot noise contributes to the half-difference map. We estimate $C_\ell^\mathrm{HS}$ and $C_\ell^\mathrm{HD}$, the half-sum and the half-difference power spectra, respectively, which are then averaged over 100 different realisations. In Fig. \ref{fig_Check_shot_noise_Pinocchio_n512}
we summarise the result. The purple squares show the averaged $C_\ell^\mathrm{HD}$ and the purple shaded regions indicate the stardard deviation of the estimated shot noise. Interestingly, we notice that the theoretical Poissonian approximation, which we indicate with the cyan, horizontal lines, is valid in every redshift bin of our analysis and in a wide range of multipoles, excluding the largest angular scales, namely $\ell \lesssim 10$, where the measured shot noise is lower than the theoretical expectation. This method shows us that we can simply subtract the Poissonian noise from the measured $C_\ell$, as in Eq. \eqref{eq_estimator_Cl}, and that the uncertainties related to this process can be quantified, and eventually included as a free parameter of the power spectrum model. Indeed, following \citet{Loureiro19} and \citet{Romanello24}, we include in the theoretical model a nuisance parameter, such that $C^{ii}_\ell\rightarrow C^{ii}_\ell +\mathcal{S}^i$. Then, $\mathcal{S}^i$ is forward modelled with a uniform prior centred in zero, and is allowed to vary within a multiple of the standard deviation estimated in the $i$-th redshift bin.
\subsection{Modelling the angular power spectrum}
The angular power spectrum is modelled from the radial projection of the spatial power spectrum, using the following relationship \citep{Padmanabhan07, Thomas11, Asorey12, Camacho19}: 
\begin{equation}
\label{eq_Cell_total}
    C_{\ell}^{ij} = \frac{2}{\pi} \int \, \mathrm{d}k \, k^2 P(k) \left[ \Psi_{\ell}^i(k) +\Psi^{i, r}_\ell(k) \right]\left[ \Psi_{\ell}^j(k) +\Psi^{j, r}_\ell(k) \right],
\end{equation}
where the projection kernel, $\Psi_{\ell}^i(k)$, describes the mapping of $k$ to $\ell$ in real space. As discussed in Sect. \ref{Sect_Selection function}, it includes the radial selection, which is summarised in $\phi^i(z)$ \citep{Asorey12}, see Eq. \eqref{eq_selection_function}, and the redshift evolution:
\begin{equation}
\label{eq_window_real_space}
    \Psi_{\ell}^i(k)=\int \, \mathrm{d}z \, b(z) \phi^i(z) D(z) j_\ell(kr(z)),
\end{equation}
where $j_\ell(x)$ is the spherical Bessel function of order $\ell$. The second term of Eq. \eqref{eq_Cell_total} accounts for the enhancement in the 2D power spectrum due to the linear large scale infall motion towards overdense regions, i.e. the Kaiser effect \citep{Padmanabhan07, Asorey12} and can be written as:
\begin{equation}
\begin{split}
\label{eq_window_redshift_space}
    \Psi^{i,r}(k)= \int \mathrm{d}z \phi^i(z) f(z) D(z) \biggl[ \frac{2\ell^2+2\ell-1}{(2\ell +3)(2\ell -1)} j_\ell(kr(z))\\ - \frac{\ell (\ell-1)}{(2\ell -1)(2\ell +1 )}j_{\ell-2}(kr(z)) -  \frac{(\ell+1)(\ell+2)}{(2\ell+1)(2\ell+3)}j_{\ell+2}(kr(z)) \biggr]. 
\end{split}
\end{equation}
Eq. \eqref{eq_window_redshift_space} derives from the assumption that the magnitudes of the peculiar velocities are small and their impact on $z_\mathrm{obs}$ is minor with respect to the radial binwidth \citep{Padmanabhan07, Thomas11}. This is even more accurate considering the peculiar velocities of galaxy clusters, whose effect on the angular power spectrum is limited only to the lower $\ell$s \citep{Romanello24}. Indeed, for $\ell \gg 0$, $\Psi^{i,r}$ tends to zero \citep{Padmanabhan07,Thomas11}, reflecting the fact that RSDs are erased during the radial integration and therefore it is not possible to resolve radial perturbations on scales smaller than the thickness of the redshift slices \citep{Padmanabhan07}. \\
\indent For the angular scales of our interest, we make use of the Limber approximation \citep{Limber53}, in order to reduce the computational cost related to the evaluation of spherical Bessel functions. Under this condition, Eq. \eqref{eq_Cell_total} reduces to:
\begin{equation}
    \label{eq_Cl_Limber}
    C_\ell^{ij}= b^i_{\mathrm{eff}} b^j_{\mathrm{eff}} \int_0^\infty \mathrm{d}z \phi^i(z)\phi^j(z) P\left(\frac{\ell+\frac{1}{2}}{r(z)}, z\right) \frac{H(z)}{r^2(z)c},
\end{equation}
where $b_\mathrm{eff}$ is computed for each redshift bin with Eq. \eqref{eq_eff_bias} and $P(k)$ is evaluated with Eq. \eqref{eq_Pk_IR_resummation}. Strictly speaking this equation is valid for small angles, namely $\ell \gg 1$, but in practice for such massive tracers it coincides with the exact computation expression already at $\ell \approx 25$ \citep{Romanello24}. As a consequence, the Limber approximation is accurate enough to reproduce the observed power spectra, given the uncertainties on the measurements of the lowest multipoles. Specifically, coloured lines in Fig. \ref{fig_Cl_pinocchio_02} are obtained with this simplified model, after the convolution with the mixing matrix (see Appendix \ref{Appendix_A}, Eq. \ref{Eq_mixing_matrix}). 


%
\section{Bayesian analysis}
\label{sect_bayesian}
We now focus on comparing the posteriors of $\xi(r)$, $w(\theta)$ and $C_\ell$. The aim is to test whether it is possible to extract consistent amount of cosmological information from these three different probes and to understand to what extent they can be complementary. In particular, we want to investigate if a strategy based on tomographic angular clustering can provide competitive results compared to the full 3D clustering, but with the advantage of not requiring cosmological assumptions to convert from redshifts to distances, during the measurement.  
\subsection{Priors and likelihood functions}
We estimate the set of cosmological parameters the clustering is more sensitive to, namely $\Omega_m$ and $\sigma_8$. For them we choose large, flat priors, between $[0.1,0.8]$ and $[0.4,1.6]$, respectively. The baryon density, $\Omega_b$, the primordial spectral index, $n_s$, and the normalised Hubble constant, $h$, are kept fixed to the values of the \textsc{Pinocchio} simulation, i.e. from \citet{Planck14}. We also choose a flat prior, between $[0,100]$, for the nonlinear damping at the BAO scale, $\Sigma_\mathrm{NL}$. In the angular power spectrum analysis we have introduced some extra shot noise parameters, $\mathcal{S}_k$, one for each independent redshift bin, which quantify the variation of the shot noise around the Poissonian value. In order not to lose constraining power, we have chosen narrow prior intervals, namely $[-3\alpha_k, +3\alpha_k]$, where $\alpha_k$ is the standard deviation of the shot noise distribution, derived from the study in Sect. \ref{sect_shot_noise_pinocchio}. Its typical value ranges from $2 \times 10^{-6}$ to $6 \times × 10^{-6}$, depending on the binwidth, $\Delta z$, and on the considered redshift bin. This choice also takes into account the fact that the Poissonian approximation holds for all the angular scales of our interest, so that the marginalised posterior distributions on $\mathcal{S}_k$ cannot deviate significantly from this value. The full list of parameters is summarised in Table \ref{table_par_pinocchio}.\\
\indent The clustering analysis is performed through Bayesian statistics, by adopting a Gaussian likelihood in the $k$-th redshift bin:
\begin{equation}
\mathcal{L}_k\propto \mathrm{exp}(-\chi^2_k/2),
\end{equation}
where: 
\begin{equation}
    \chi^2_k=\sum_{a=1}^{N_k} \sum_{b=1}^{N_k} (\mu^{d}_a-\mu^m_a)_{(k)} \hat{C}^{-1}_{abkk} (\mu^{d}_b-\mu^m_b)_{(k)},
\end{equation}
where $N_k$ is the number of bins in the $k$-th redshift range, $a$ and $b$ refer to the bin involved, $\mu$ is the clustering statistic, i.e. the correlation function or the power spectrum, with superscripts $d$ and $m$ if the quantity is measured from the data or computed with the model, respectively. $\hat{C}^{-1}_{abkk}$ is the inverse of the numerical covariance matrix in the $k$-th redshift bin, estimated directly from 1000 mocks, using $\xi(r)$, $w(\theta)$ or $C_\ell$ in Eq. \eqref{eq_numerical_covariance}. As already discussed, from an inspection of the blocks outside the diagonal we conclude that the cross-correlation between different redshift bins is consistent with zero, therefore the total likelihood simply reduces to the product of the individual likelihoods, $\mathcal{L}_k$.
\begin{table*}[h!]
\caption{Parameters and priors considered in the cosmological analyses. The square brackets delimit the intervals of the uniform priors, while the fixed numbers indicate which of the cosmological parameters are assumed to be constant. The extra shot noise term, $\mathcal{S}_k$, is a free parameter only for the angular power spectrum analysis. The normalised Hubble constant, $h$, is defined as $H_0/(100 \, \mathrm{km \, s^{-1} \, Mpc ^{-1}})$.}        
\label{table_par_pinocchio}      
\centering
\small
\begin{tabular}{c c c}     
\hline\hline       
Parameter & Description & Prior \\ 
\hline                    
    $\Omega_{\mathrm{m}}$  & Total matter density parameter & $[0.1, 0.8]$ \\
    
    $\sigma_8$  & Amplitude of the power spectrum on the scale of 8 Mpc $h^{-1}$ & $[0.4, 1.6]$ \\
    
    $\Omega_\mathrm{b}$  & Baryon density parameter & $0.48254$ \\

    $n_\mathrm{s}$  & Primordial spectral index & $0.96$ \\

    $h$  & Normalised Hubble constant & $0.6777$ \\
    
    $\Sigma_\mathrm{NL}^k$  & Nonlinear damping at BAO scale & $\;[0, 100]$ \\
    
    $\mathcal{S}_k$  & Extra shot noise parameter & $\;[-3\alpha_k, +3\alpha_k]$ \\
    
\hline                  
\end{tabular}
\end{table*}
\subsection{Results}
\begin{figure*}[htpb]
\centering
\includegraphics[width=0.8\hsize]{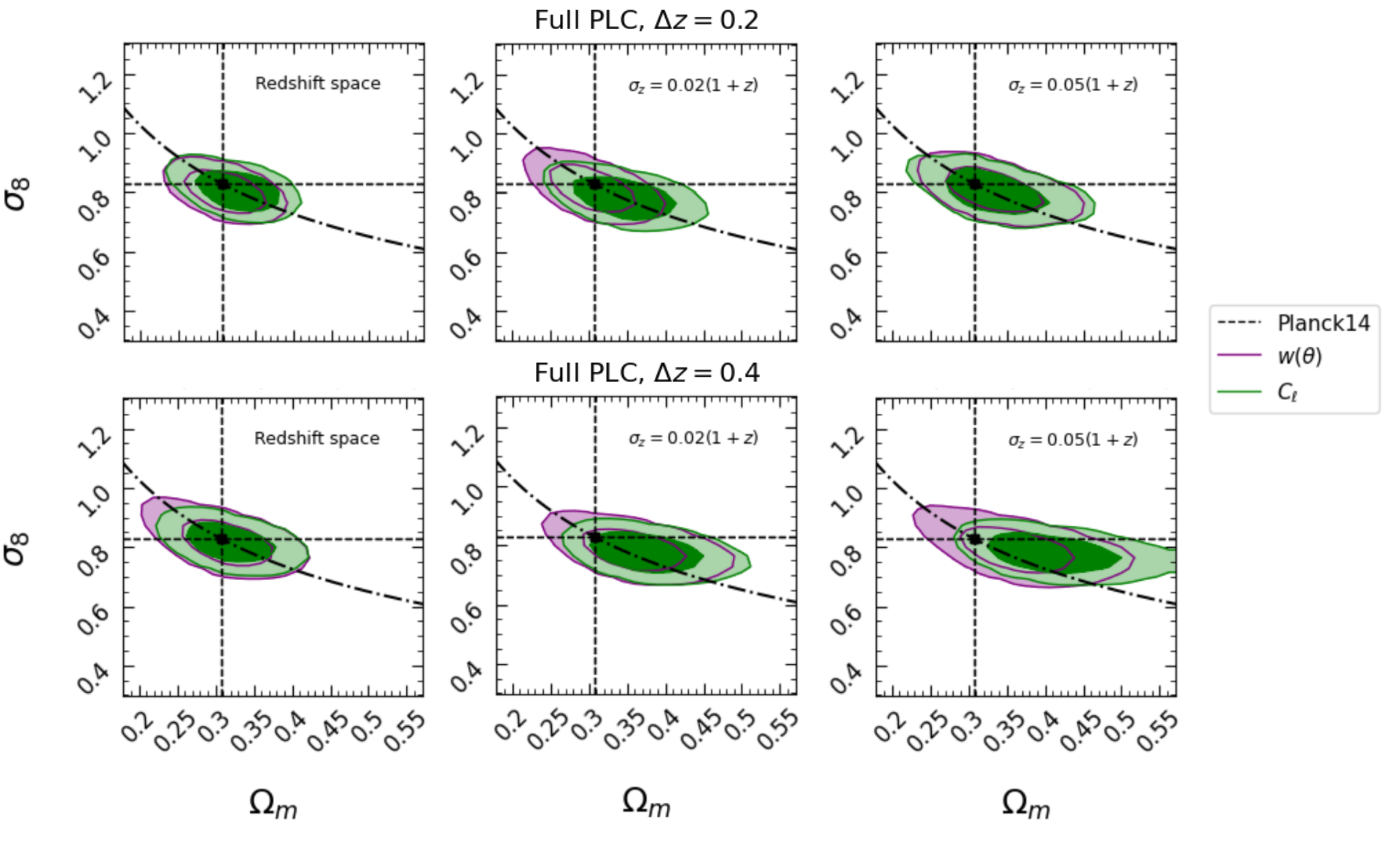}
\caption{The marginalised posterior distributions in the $\Omega_m-\sigma_8$ plane, with 68\% and 95\% confidence intervals, for redshift and photo-$z$ space, using a bindwidth of $\Delta z=0.2$ and $\Delta z=0.4$, in the upper and lower panels, respectively. Contours for $w(\theta)$ and $C_\ell$. The dashed lines indicate the $\Omega_m$ and $\sigma_8$ values of the \textsc{Pinocchio} simulation, while the dot-dashed line shows the predicted $S_8$ degeneracy curve.}
\label{fig_Corner_2D}
\end{figure*}

In this section we present the results of the cosmological analyses of $\xi(r)$, $w(\theta)$ and $C_\ell$. We sampled the posterior distribution with a Monte Carlo Markov Chain (MCMC), assuming different redshift bins as statistically independent. For all the probes considered, the cosmological parameters of the simulations, corresponding to \citet{Planck14}, are correctly recovered, falling within the range at 95\% of the corner plot in the $\Omega_m-\sigma_8$ plane. Systematic effects causing a slight shift of the marginalised posterior distribution can rise due to nonlinearities or to the inaccuracy of the halo bias model. Indeed, the theoretical prediction provided by \citet{Tinker10} overestimates the numerical bias by about 5\%, for low masses and redshifts, see Fig. 2 in \citet{Fumagalli_counts}. On the other hand, at high masses and redshifts, there is an underestimation of about 10\%. However, these scales are excluded from the redshift cut of our sample.\\
\indent In Fig. \ref{fig_Corner_2D} we summarise the result of the angular tomographic analysis. First, we note that $w(\theta)$ and $C_\ell$ produce largely overlapping results, giving us the same cosmological information. This is due to the fact that our catalogue has a large coverage of the sky, allowing us to investigate a wide range of different wavelengths. The relative shift of their posteriors, for example, for $\Delta z=0.2$ and $\sigma_z=0.02(1+z)$ or $\Delta z=0.4$ and $\sigma_z=0.05(1+z)$, can be interpreted as an effect of the variance of the specific PLC, and it is not found when instead of measuring on a single mock we consider the average measure from the full set of 1000 \textsc{Pinocchio} simulations. \\
\indent Second, by widening the binwidth from $\Delta z=0.2$ to $\Delta z=0.4$, the marginalised posterior distributions of $w(\theta)$ and $C_\ell$ suffer a broadening on both $\Omega_m$ and $\sigma_8$. For example, the constraints from the angular correlation function slightly change from $0.31^{+0.03}_{-0.03}$ to $0.32^{+0.04}_{-0.04}$, and from $0.80^{+0.03}_{-0.03}$ to $0.81^{+0.04}_{-0.03}$, respectively. In fact, with a binwidth of $\Delta z=0.4$ we have both more statistics and less shot noise, as we include a larger number of dark matter haloes. However, the effective bias is considered as a constant, within a given redshift bin, to be taken out of the integrals in Eqs. \eqref{eq_wtheta_model} and \eqref{eq_Cl_Limber}. This may have systematic consequences for the parameters estimation. Furthermore, the 2D clustering is modified by projection effects that aggregate structures that are not physically bound. In addition, the deeper radial projection erases the BAO peak, because its physical size ends up in a wider range of angular scales. This highlights the importance of defining an appropriate tomographic strategy in clustering studies.\\
\begin{figure*}[htpb]
\centering
\includegraphics[width=0.8\hsize]{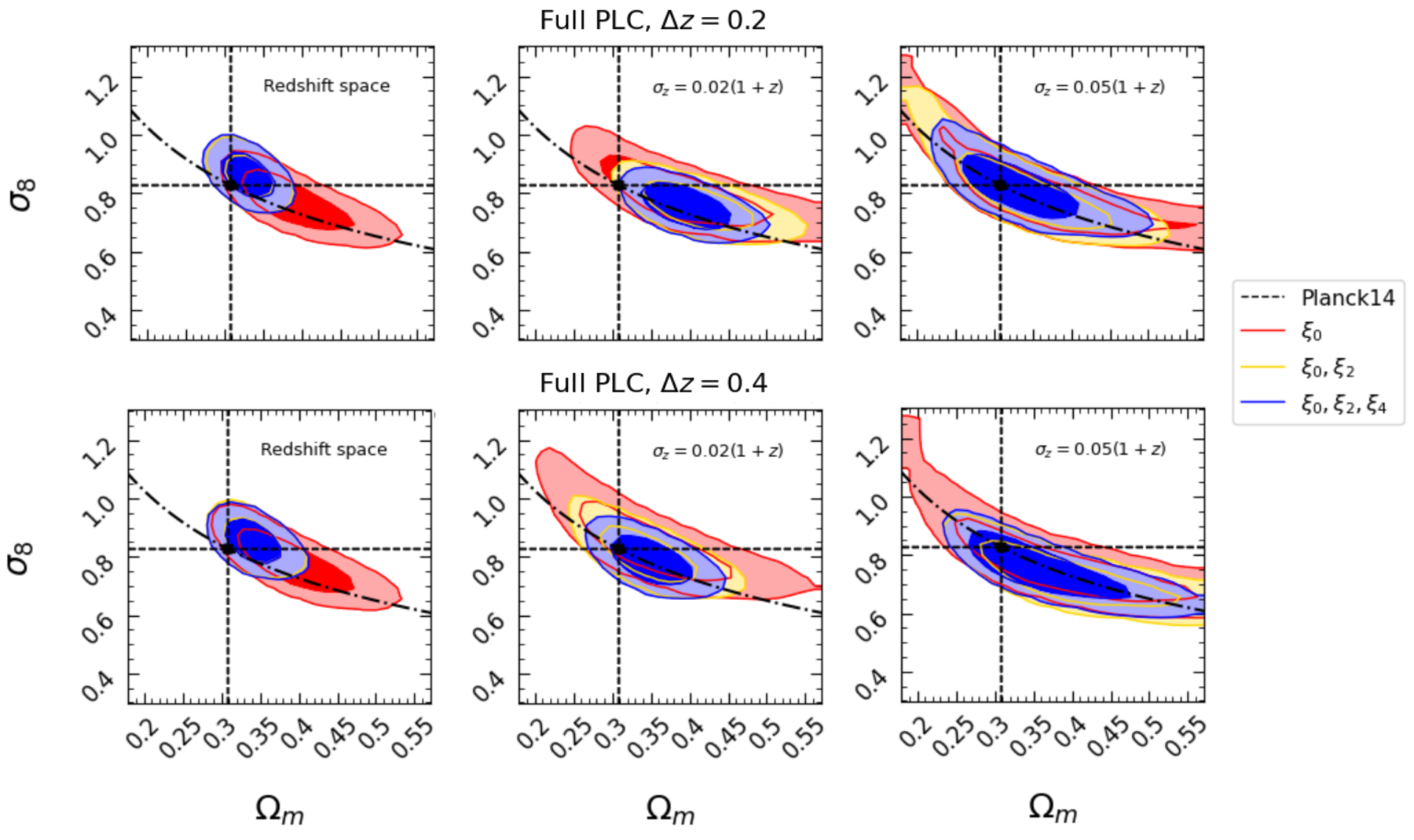}
\caption{The marginalised posterior distributions in the $\Omega_m-\sigma_8$ plane, with 68\% and 95\% confidence intervals, for redshift and photo-$z$ space, using a bindwidth of $\Delta z=0.2$ and $\Delta z=0.4$, in the upper and lower panels, respectively. Contours for the different multipoles of $\xi(r)$. The dashed lines indicate the $\Omega_m$ and $\sigma_8$ values of the \textsc{Pinocchio} simulation, while the dot-dashed line shows the predicted $S_8$ degeneracy curve.}
\label{fig_Corner_3D}
\end{figure*}
\indent The introduction of photometric errors, on the other hand, determines a general reduction of the clustering signal, resulting in a loss of constraining power. The posterior distributions given by $w(\theta)$ broaden from $0.32^{+0.03}_{-0.03}$ to $0.33^{+0.04}_{-0.04}$, and from $0.80^{+0.03}_{-0.03}$ to $0.81^{+0.03}_{-0.03}$, if $\Delta z=0.2$, while they widen from $0.32^{+0.04}_{-0.04}$ to $0.36^{+0.05}_{-0.06}$, and from $0.81^{+0.04}_{-0.03}$ to $0.79^{+0.05}_{-0.03}$, if $\Delta z=0.4$, respectively. These results show us that the constraints on $\sigma_8$ are generally more stable than the ones on $\Omega_m$.\\
\indent As we can see in Table \ref{table_par_pinocchio}, we have also included some extra shot noise terms for the angular power spectrum. While in general we have verified that these shot noise parameters are characterised by a flat zero mean posterior, we notice that the $\Omega_m-\sigma_8$ corner plots, obtained by including these additional free parameters, are slightly wider and better centred.\\
\indent  In Fig. \ref{fig_Corner_3D} we can see the posterior distributions derived from the 3D correlation function. As in many studies based on real observations, we start from the analysis of the monopole, since it is more stable and less sensitive to the purity of the cluster sample. Then, we progressively take into account the contribution of the quadrupole and the hexadecapole. Focusing on $\xi_0(r)$, we found that enlarging the redshift shell from $\Delta z=0.2$ to $\Delta z=0.4$ does not change significantly the redshift-space marginalised distribution, with a maximum shift of $0.01$ in the standard deviations of both $\Omega_m$ and $\sigma_8$. This result should be interpreted with the fact that in the 3D correlation function the redshift information is not integrated out, thus it contributes to the clustering signal, with only a minor degradation due to the assumption of a constant effective bias in each tomographic bin. \\
\indent On the other hand, we note that $\xi_0(r)$, in contrast to $w(\theta)$ and $C_\ell$, exhibits a significant widening of the confidence intervals as the photometric error increases. While, from the point of view of the model, the damping of the power spectrum in Eq. \eqref{eq_xi_damped_sereno} is fully equivalent to the introduction of Gaussian errors in the $\mathrm{d}N/\mathrm{d}z$, as shown in \citet{Hutsi10}, it is the measurement that is mainly altered, since the photometric error significantly modifies the radial component of the clustering, leaving the angular part unaffected. For example, while the 3D correlation function suffers from the complete erosion of the BAO peak, this is still visible in $w(\theta)$, which only experiences a reduction in the signal due to differences in the radial distribution within a given photometric redshift interval (see Fig. \ref{fig_redshift_distribution_pinocchio}). A second consequence is that, in photo-$z$ space, the parameter $\Sigma_\mathrm{NL}$ is not constrained, due to the smearing that affects the BAO peak. Therefore it behaves as a merely computational expensive complication of the model.\\
\indent Furthermore, the constraints are better centred on the true cosmology of the simulation as the photometric errors increase. This is because we start from the assumption of a linear model for $P(k)$, with an infrared (IR) resummation at the BAO peak. As we can see in Fig. \ref{fig_3D_correlation_pinocchio_02}, this approximation does not hold at low redshift and small scales, because of deviations due to the FoG effect and the nonlinear coupling between velocity and density fields \citep[see e.g.][]{TNS10}. This could result in a systematic on the estimate of cosmological parameters. Conversely in photo-$z$ space, the effect of nonlinearities is diluted by the redshift perturbation, in a similar way to the redshift projection of 2D clustering. This makes it possible to still considering a simple linear power spectrum, avoiding complications in the modelling of the clustering signal. \\
\indent As evident from Fig. \ref{fig_Corner_3D}, the inclusion of multipoles leads, in general, to a shrinking of the contours along the $S_8\equiv\sigma_8\sqrt{\Omega_m/0.3}$ degeneration line. We should notice, in particular, that in redshift space the clustering signal comes from the monopole and the quadrupole only, while the contribution of the hexadecapole is negligible, as we can see in Fig. \ref{fig_3D_correlation_pinocchio_02}, being consistent to zero for $r>20$ Mpc $h^{-1}$ in every bin of our analysis. On the other hand, in photo-$z$ space, the inclusion of $\xi_2(r)$ and $\xi_4(r)$ is relevant. By way of example, for $\Delta z=0.4$, the constraints on $\Omega_m$ and $\sigma_8$ change from $0.35^{+0.08}_{-0.06}$ and $0.83^{+0.10}_{-0.07}$, to $0.34^{+0.04}_{-0.04}$ and $0.80^{+0.06}_{-0.05}$, and finally to $0.35^{+0.03}_{-0.03}$ and $0.78^{+0.04}_{-0.03}$, as we add the even multipoles.\\
\indent In Fig. \ref{fig_S8}, in order to summarise the cosmological results and to understand what contribution cluster clustering can make, we focus on the structure growth parameter, $S_8$. It is particularly relevant in cosmology, since various measurements in the local Universe, e.g. cosmic shear, galaxy clustering, galaxy and galaxy cluster number counts, have shown a $2-3\sigma$ tension with the high-$z$ value inferred by studying the CMB anisotropies from Planck \citep[see e.g.][]{Bocquet19, Asgari21, Heymans21, Abbott22, Amon22}, which represents a potential stress for the current $\Lambda$CDM model. \\
\indent For the sake of simplicity, we compare only the two-point correlation function in both its variants, i.e. $w(\theta)$ and $\xi(r)$. These two probes lie in configuration space. Moreover, they are built from the same random catalogue and therefore are subject to common systematics. While in redshift space their competitiveness essentially depends on the width of the analysed redshift bins, with a slight decentring of $\xi_0(r)$ due to nonlinearities at small scales, we note that in the presence of a large photometric error, such as $0.05(1+z)$, $w(\theta)$ is always favoured over $\xi_0(r)$, with $0.85^{+0.03}_{-0.03}$ versus $0.87^{+0.06}_{-0.04}$ and $0.86^{+0.04}_{-0.03}$, versus $0.86^{+0.07}_{-0.05}$, for $\Delta z=0.2$ and $\Delta z=0.4$, respectively. In fact, the latter requires the inclusion of the multipoles to produce results comparable or superior to the 2D clustering, depending on $\Delta z$. In this case, more care should be taken to avoid systematics due to nonlinearities not included in the model, at scales $r<20$  Mpc $h^{-1}$. The intermediate case, with $\sigma_z=0.02(1+z)$, shows that the 2D correlation function is clearly preferred for $\Delta z=0.2$, while it is slightly disfavoured for $\Delta z=0.4$, due to the lack of the redshift information. This fact confirms the possible advantages and the competitiveness of the angular tomographic clustering within photometric redshift surveys. In particular, the potential and limitations of this approach have already been tested in the angular clustering analysis of the KiDS-DR3 cluster catalogue, discussed in \citet{Romanello24}, where the constraint on $S_8$ are mildly more stringent than the corresponding 3D study \citep{Lesci22}.
\begin{figure*}[htbp]
\centering
\includegraphics[width=\hsize]{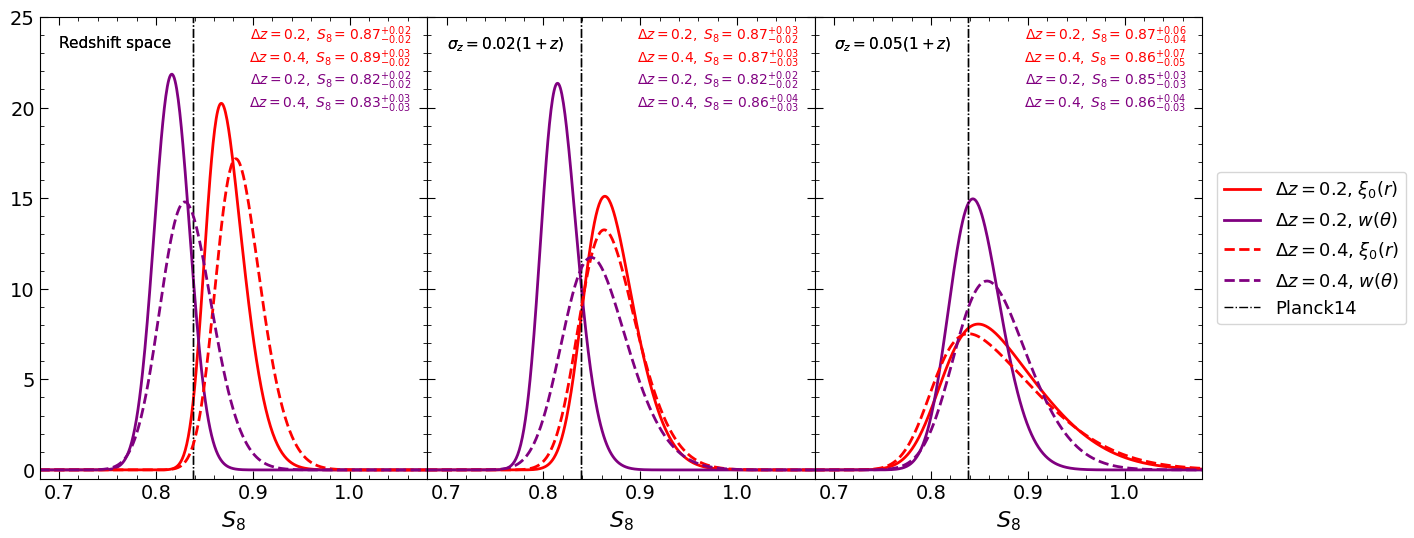}
\caption{Summary plot for the parameter $S_8$, in redshift and photo-$z$ space with $\sigma_z=0.02(1+z)$ and $\sigma_z=0.05(1+z)$, panels from left to right, for $w(\theta)$ and $\xi_0(r)$. The solid lines refer to $\Delta z=0.2$, the dashed lines are for $\Delta z=0.4$. The dot-dashed line indicates the $S_8$ value of the \textsc{Pinocchio} simulation.}
\label{fig_S8}
\end{figure*}
\section{Conclusions}
\label{sect_conclusions}
In this paper, we have analysed the clustering properties of dark matter haloes from the \textsc{Pinocchio} simulation, in preparation for subsequent work to be carried out on forthcoming galaxy cluster catalogues, from Stage-III and Stage-IV photometric surveys. We used a PLC with an angular size of 60 degrees, thus covering a circular area of 10313 deg$^2$, namely a quarter of the sky. We selected dark matter haloes with a virial mass greater than $10^{14} M_\odot \, h^{-1}$ and redshift in the range $0.2<z<1.0$. The mass selection reproduces to a good approximation what is expected for future cluster catalogues identified in the Stage-IV photometric surveys, while the selection in $z$ excludes the low redshift regime, where it will be difficult to perform a robust lensing analysis for the mass calibration of real galaxy clusters, and the high redshift regime, where the clustering signal is completely dominated by shot noise. \\
\indent Our study used a tomographic approach adopting the binwidths $\Delta z=0.2-0.4$, i.e. 4 or 2 redshift bins, to compare 3D and 2D clustering. The former is analysed by means of the 3D two-point correlation function, while for the latter we made use of the angular correlation function and its spherical harmonic counterpart, the angular power spectrum. For each cosmological probe, we estimated the numerical covariance matrix from a set of 1000 mock catalogues, in both redshift space and photo-$z$ space. For this purpose, we introduced Gaussian errors on the halo redshifts, with zero mean and standard deviation given by $\sigma_z=\sigma_{z,0}(1+z)$, with the typical $\sigma_{z,0}$ of photometric redshift surveys, namely $0.02$ and $0.05$. \\
\indent We used a linear model for the power spectrum, including the IR-resummation, which characterises nonlinear damping phenomena at the BAO scale. The 3D correlation function is obtained by Fourier transforming the corresponding $P(k)$, computing the effective bias of dark matter haloes by assuming the model presented in \citet{Tinker10}, for the halo bias, and the parameterisation of \citet{Despali16}, for the halo mass function. The angular correlation function and power spectrum are modelled from the radial projection of the corresponding 3D correlation function and power spectrum, respectively, with a selection kernel that takes into account the normalised redshift distribution of haloes. RSDs are modelled in the Kaiser limit, for the correlation functions, while they are not considered in the $C_\ell$, as they only slightly affect the largest angular scales, being fully subdominant relative to the measurement uncertainties. Indeed, we found that they have only a minor effect, both on $\xi_0(r)$ and $w(\theta)$, quantified in a boost factor of 1.1 - 1.2, due to the high bias of the tracers. On the other hand, photometric errors play the major role in suppressing the clustering signal and the numerical errors. In particular, they modify the small-scale slope of the 3D correlation function monopole, and completely erase the BAO features, which conversely remain visible in $w(\theta)$. Moreover, photometric errors determine a substantial increase in the contribution of the quadrupole and the hexadecapole. In particular, $\xi_2(r)$ changes its sign, moving from redshift space to photo-$z$ space, while $\xi_4(r)$, which is consistent with zero in redshift space, becomes dominant for $\sigma_z \gtrsim 0.02(1+z)$. For the angular power spectrum, we verified that the Limber approximation and the Poissonian assumption for the shot noise hold over the full redshift interval and angular scales considered in our analysis. \\
\indent Finally, we perform a Bayesian MCMC analysis to constrain fundamental cosmological parameters such as $\Omega_m$ and $\sigma_8$. We found that despite the fact that all cosmological probes are able to accurately estimate the cosmological parameters of the simulations, the posterior distributions obtained from the 3D correlation function degrade more in the presence of photometric errors than from its angular counterparts. Indeed, in the presence of large photometric errors, $\xi_0(r)$ produces wider constraints on the structure growth parameter, $S_8$, with respect to $w(\theta)$. Conversely, by exploiting also the redshift infomation, it is less affected by the specific tomographic redshift binning strategy adopted, but more sensitive to nonlinearities that characterise the small-scale clustering. \\
\indent The inclusion of the two-point correlation function multipoles determines a collapse of the $\Omega_m-\sigma_8$ contours along the $S_8$ degeneracy curve, which is more relevant in photo-$z$ space, while the contribution of $\xi_4(r)$ is negligible in redshift space. \\
\indent On the other hand, $w(\theta)$ and $C_\ell$ provide the same cosmological information, and unbiased cosmological constraints, which in turn complement that obtained from the 3D analysis, with the advantage of being able to rely on a linear model for $P(k)$, since nonlinearities are diluted by projection effects. Moreover, in the presence of photometric errors, for $\sigma_{z,0} \gtrsim 0.02$, the angular clustering shows its full potential as a cosmological probe, with respect to the traditional 3D study, with the advantage of not requiring a fiducial cosmology to convert redshifts into distances in the measurements. This comparison was already performed in \citet{Romanello24}, where we successfully applied the models developed in this paper to the AMICO KiDS-DR3 galaxy cluster catalogue, finding consistent results with respect to the $\xi_0(r)$ analysis, with slightly more stringent constraints on $S_8$. This highlights the significance of extending the analysis of 2D clustering to other cosmological parameters, e.g. to study the degeneracy in the $w_0 - w_a$ dynamical dark energy plane, and to the galaxy cluster catalogues that will be available from ongoing Stage-III and Stage-IV photometric surveys, such as the forthcoming data releases of KiDS and \textit{Euclid}.

\begin{acknowledgements}
     We acknowledge the financial contribution from the grant PRIN-MUR 2022 20227RNLY3 “The concordance cosmological model: stress-tests with galaxy clusters” supported by Next Generation EU and from the grants ASI n.2018-23-HH.0 and n. 2024-10-HH.0 “Attività scientifiche per la missione Euclid – fase E”, and the use of computational resources from the parallel computing cluster of the Open Physics Hub (\url{https://site.unibo.it/openphysicshub/en}) at the Physics and Astronomy Department in Bologna. We would like to thank P. Monaco for the production of the \textsc{Pinocchio} mocks and R. Zangarelli for the constructive suggestions that improved this work.
\end{acknowledgements}

%
\bibliographystyle{aa} 
\bibliography{bibliografia} 
%

\begin{appendix} 
\section{Reducing the sky coverage}
\label{Appendix_A}
\begin{figure*}[htbp]
\centering
\includegraphics[width=0.6\textwidth]{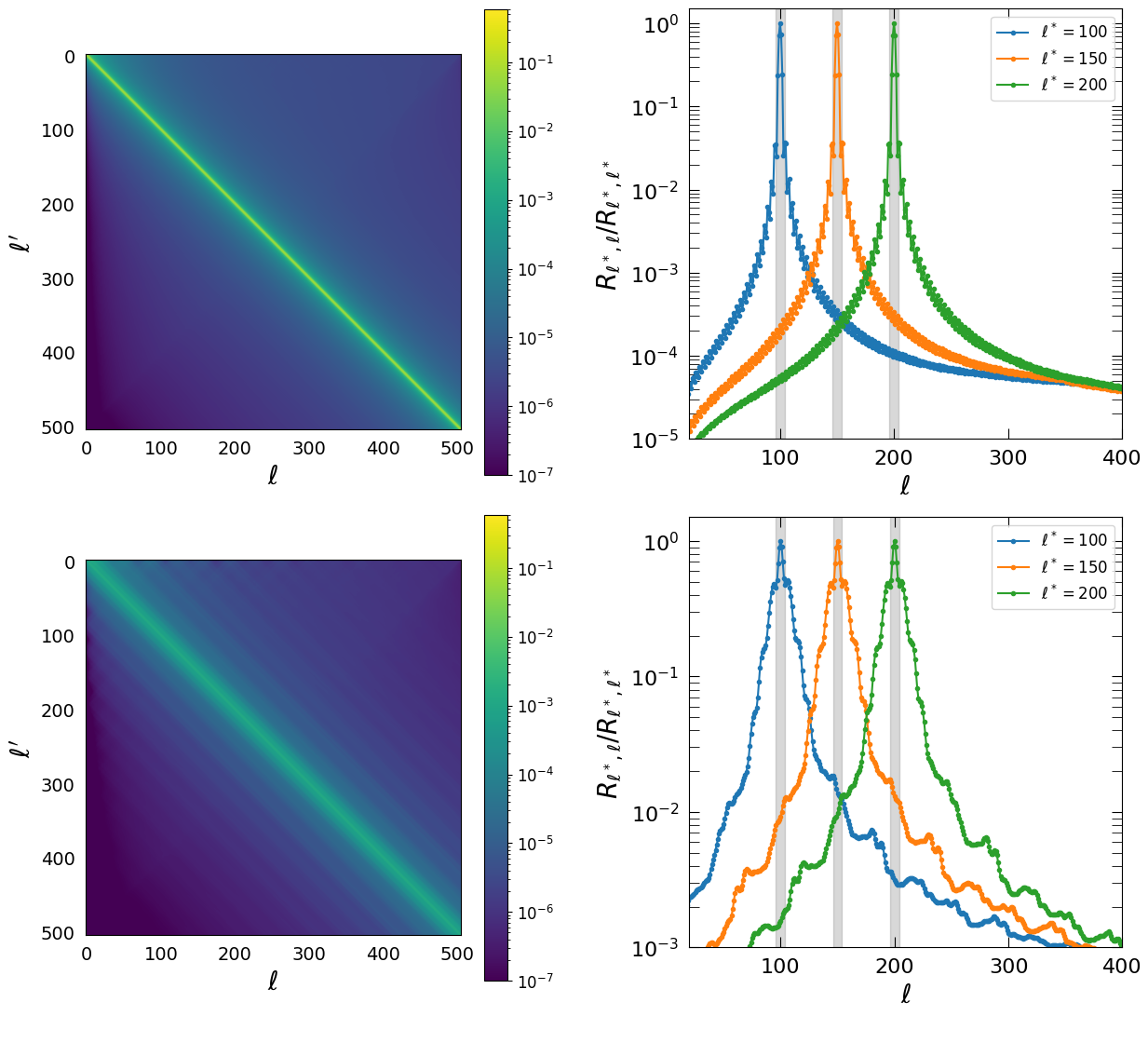}
\caption{Left panels: colour plots of the mixing matrix, $R_{\ell \ell'}$, computed up to $\ell = 500$ with the public code \textsc{NaMaster} \citep{Alonso19}. Right panels: plots of $R_{\ell \ell^*}/R_{\ell^* \ell^*}$, centred in three different multipoles, at $\ell=100$ (blue), 150 (orange) and 200 (green). They quantify the mode-mode coupling related to the partial sky coverage, reflecting the blurring of the mixing matrix along the diagonal. The grey shaded regions indicate the multipole range of $\Delta \ell$ in which we bin our measurements. The top panels are computed from the 10313 deg$^2$ region covered by the full \textsc{Pinocchio} footprint mask, while at the bottom we focus on the 1004 deg$^2$ subregion.}
\label{fig_Mixing_matrix_pinocchio}
\end{figure*} 
\indent In partial-sky surveys, spherical harmonics no longer provide a complete orthonormal basis for the expansion of the angular overdensities \citep{Camacho19}. The masked density field is related to its full-sky counterpart through a binary mask function, $\Tilde{\delta}_{\mathrm{h}}(\boldsymbol{\hat{n}})=M(\boldsymbol{\hat{n}})\delta_{\mathrm{h}}(\boldsymbol{\hat{n}})$, which can be considered as an angular window function. Its effect propagates into the power spectrum evaluation, in the form of a coupling between different multipoles, which becomes progressively more important for lower sky coverages and survey fragmentation. In particular, the measured power spectrum at multipole $\ell$ depends on an underlying range of multipoles $\ell'$ \citep{Blake07}. The power transfer between different multipoles is described by the mixing matrix, $R_{\ell \ell '}$, which depends only on the geometry of the angular mask. It reduces to the identity matrix in full-sky surveys and can be expressed in terms of the Wigner $3j$ symbols:  
\begin{equation}
    \label{eq_mixing_matrix_wigner}
    R_{\ell \ell' }=\frac{2 \ell ' +1}{4 \pi}\sum_{\ell ''} (2\ell '' +1 ) W_{\ell '' }\begin{pmatrix} \ell & \ell ' & \ell '' \\0 & 0 & 0 \end{pmatrix}  ^2, 
\end{equation}
where:
\begin{equation}
    W_\ell= \sum_{m=-\ell}^{+\ell} \frac{|I_{\ell m}|^2}{(2 \ell +1)}. 
\end{equation}
Here, $I_{\ell m}$ represents the spherical harmonic coefficient of the mask, which can be approximated in a pixelated sky as: 
\begin{equation}
    I_{\ell m} \simeq \sum^{N_\mathrm{pix}}_p Y^{*}_{\ell m}(\theta_p, \varphi_p)\Delta \Omega_p. 
\end{equation}
With the mixing matrix we can link the ensemble average of the measured power spectrum with the theoretical one through \citep{Balaguera18}: 
\begin{equation}
\label{Eq_mixing_matrix}
    \langle K^{ij}_\ell \rangle =\frac{1}{f_{\mathrm{sky}}} \sum_{\ell'} R_{\ell \ell'} C_\ell.
\end{equation}
We estimate the mixing matrix of the footprint masks using the publicly available code \textsc{NaMaster}\footnote{\url{https://github.com/LSSTDESC/NaMaster}} \citep{Alonso19}, which provides a general framework for the pseudo-$C_\ell$ analysis.\\
\indent To better reproduce the observational conditions of photometric surveys, which often scan separated parts of the sky, we produced an angular mask composed of three different patches, two near the equator (\textit{P1} and \textit{P2}), and another further south, (\textit{P3}). Their angular limits are given respectively by: 
\begin{itemize}
    \item \textit{P1}: RA $\in [-60, -30]$, Dec $\in [-10, 0]$; 
    \item \textit{P2}: RA $\in [0, 30]$, Dec $\in [4, 10]$; 
    \item \textit{P3}: RA $\in [-30, 30]$, Dec $\in [-40, -30]$.
\end{itemize}
The final effective area is equal to $1004$ deg$^2$, which simulates the expected sky coverage of the cluster catalogue of KiDS-DR4, for which we plan to carry out a subsequent in-depth study. \\
\indent Reducing the sky coverage has important consequences on our analysis. First, it increases the angular auto-correlation, with a blurring of the correlation matrix both in configuration and in spherical harmonic space. Indeed, at the level of the angular power spectrum, we find that the mixing matrix calculated over an area of 1004 deg$^2$ is more blurred than the one computed over the full PLC, as we can see in Fig. \ref{fig_Mixing_matrix_pinocchio}, with the convolution function $R_{\ell^* \ell'}$. The grey shaded regions highlight the $\Delta \ell=8$ bandwidth, which allows us to include almost all of the clustering signal, reducing the power transfer between different $\ell$ bands due to the off-diagonal terms, while with the reduced sky coverage, the same band of $\Delta \ell=8$ identifies only 50\% of the transferred power. \\
\indent Reducing the survey area, on the other hand, does not change the shot noise, since both $\Omega_\mathrm{sky}$ and $N_\mathrm{h}$ decrease in the same way. However, a smaller range of wavelengths can be investigated within the reduced volume. This fact suggests that it is possible to exploit $C_\ell$ and $w(\theta)$ as complementary sources of cosmological information and specifically, to use their comparison to assess possible model systematics.

\section{Analytical modelling of the covariance matrix}
\label{sect_model_covariance}
\indent A fundamental aspect of cluster clustering cosmology is the correct evaluation of measurement errors. Indeed, biases in the determination of cosmological parameters can be present, if all significant observational systematic uncertainties are not properly taken into account.\\
\indent The estimation of covariance matrices, which summarise the set of uncertainties associated with the observables, is therefore a fundamental challenge for the future. Analytical covariance matrix calculations are probably the best solution, as they do not require large computational resources. In addition, they are noise-free and naturally include the dependence on cosmological parameters \citep{Fumagalli_clustering}. However, the accuracy of analytical models might be limited due to the complex physics of structure formation, which introduces nonlinearities and non-Gaussianities in clustering, and by the necessity of appropriate models for shot noise and RSDs  \citep{Manera13, Fumagalli_clustering}. These and other observational effects, such as angular masks and redshift uncertainties, can lead to systematic theoretical errors, and therefore their modelling needs to be validated against simulations in any case \citep{Avila18, Fumagalli_covariance, Fumagalli_clustering}.\\
\indent In this section, we will test some simple theoretical models for the covariance matrix of the angular two-point correlation function and power spectrum. The validation of semi-analytic models for the covariance of the 2D clustering is beyond the scope of this paper, while for the 3D two-point correlation function we rely on the results presented in the works by \citet{Fumagalli_covariance, Fumagalli_clustering}.
\subsection{Covariance matrix of \texorpdfstring{$C_\ell$} {}}
\label{sec_covariance_cl}
\begin{figure*}[htbp]
\centering
\includegraphics[width=0.8\textwidth]{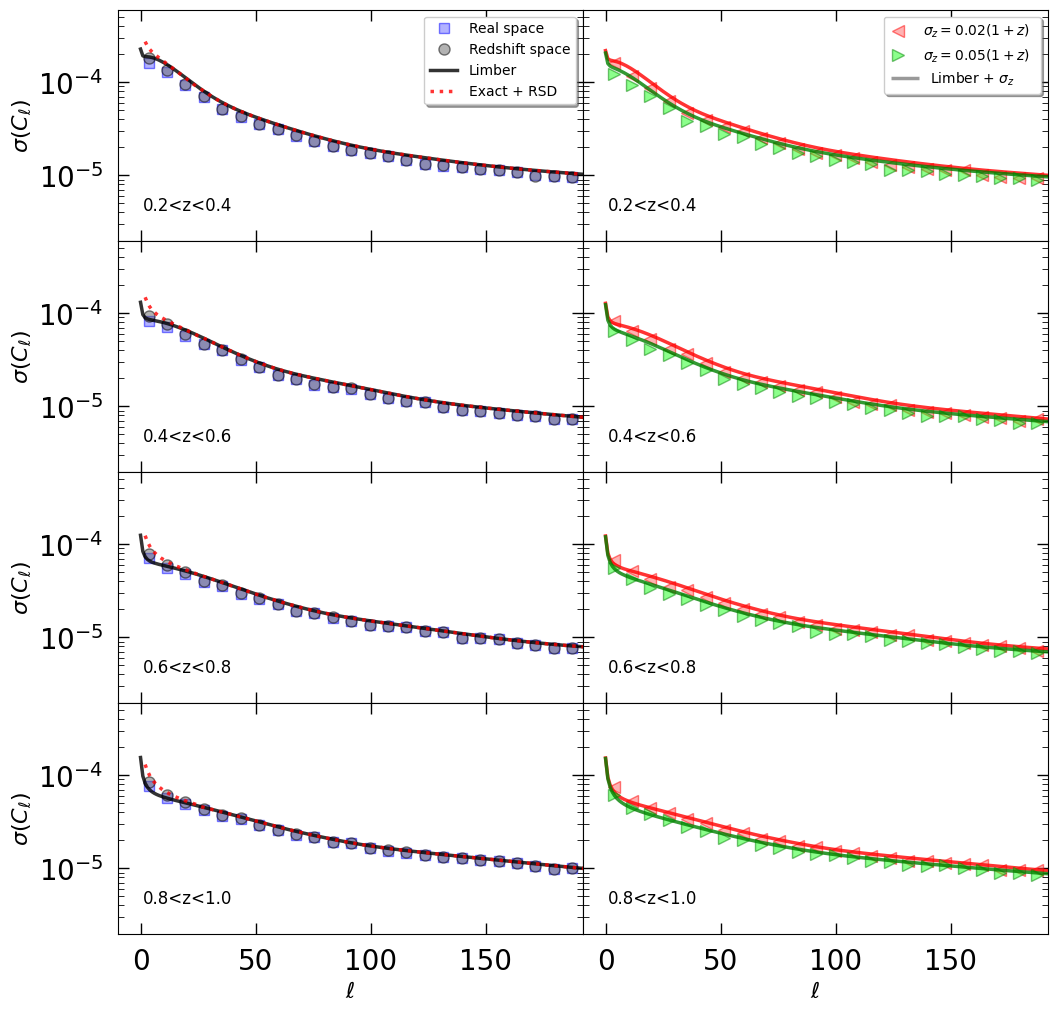}
\caption{Errors in the angular power spectrum, derived as the square root of diagonal elements of $\mathrm{Cov}_{\ell \ell'}$, in real (blue squares), redshift (with $\sigma_z=0$, black circles) and in photo-$z$ space, with $\sigma_z=0.02$, (red triangles) and $\sigma_z=0.05$ (green triangles), from the numerical covariance matrix measured from 1000 mocks. The lines show the results obtained with the linear power spectrum in the Limber approximation (solid black line), with the exact integration given by Eq. \eqref{eq_Cell_total}, which also includes the modelling of the RSDs (dashed red line) and with the damped linear power spectrum in the Limber approximation (solid red and green lines).}
\label{fig_Cl_diagonal_cov02}
\end{figure*}
\begin{figure*}[htbp]
\centering
\includegraphics[width=0.8\textwidth]{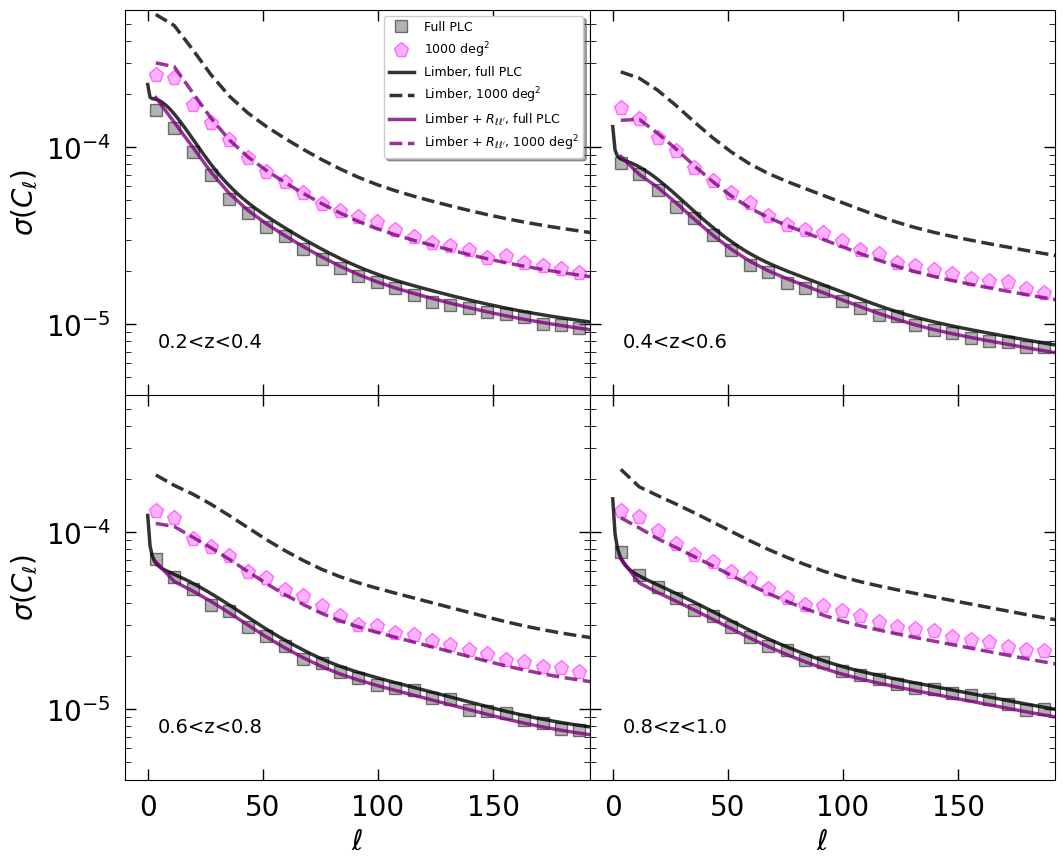}
\caption{Comparison between diagonal and non-diagonal errors on the real-space angular power spectrum, $C_\ell$, in different redshift bins. Grey squares and magenta pentagons are measured from the set of mocks, with the full PLCs sky coverage and with its 1000 deg$^2$ subregions, respectively. 
The black lines are computed through the Limber approximation, with the diagonal $1/\sqrt{f_\mathrm{sky}}$ boost factor. The purple lines account for the mode-mode coupling induced by the mixing matrix, $R_{\ell,\ell'}$. Solid lines refer to the full PLCs sky coverage, while dashed lines refer to 1000 deg$^2$ subregions.}
\label{fig_Cl_diagonal_mixing_cov02}
\end{figure*}

Concerning angular clustering, several works in the literature considered internal subsampling error estimates, such as bootstrap or jackknife \citep[e.g.][]{Paech17, Balaguera18}. These methods are quite accurate for 2D clustering, because the radial projection alleviates the tensions that emerge in the 3D correlation \citep{Norberg09}, depending on the scales involved, ensuring a good agreement with theoretical models \citep{Crocce11}. \\
\indent In the linear regime we can assume a Gaussian distribution for the $a_{\ell m}$s. As a consequence of the 
Gaussian theory, covariance matrices are limited to diagonal terms \citep{Dodelson03}. This analytical estimation has been used in several works \citep{Blake07, Thomas11, Balaguera18} and can be written as follows:
\begin{equation}
    \label{eq_covariance_Cl}
    \mathrm{Cov_{\ell \ell'}}=\frac{2}{(2\ell +1)\Delta \ell f_\mathrm{sky}}\left(C_\ell+\frac{\Omega_\mathrm{sky}}{N_\mathrm{h}}\right)^2 \delta^K_{\ell \ell'},
\end{equation}
where the Kronecker delta, $\delta^K_{\ell \ell'}$, ensures the diagonality of the covariance matrix. It consists of an average over $2\ell + 1$ different realisations, one for each $m$ in a given multipole, with the cosmic variance term, $C_\ell$, and the shot noise term, $\Omega_\mathrm{sky}/N_\mathrm{h}$, that depends on the number of haloes in the selected redshift bin. The partial sky coverage enters in Eq. \eqref{eq_covariance_Cl} as a boost factor $1/f_\mathrm{sky}$, though the covariance matrix remains diagonal. This is accurate enough because, as we have already discussed, the off-diagonal elements of the correlation matrix become negligible when averaging in the $\ell$ bands and consequently dividing for their size $\Delta \ell$. Eq. \eqref{eq_covariance_Cl} depends on the theoretical power spectrum $C_\ell$, which includes the modelling of observational effects, such as photometric errors and RSDs, according to Eqs. \eqref{eq_selection_function} and \eqref{eq_Cell_total}, respectively. In Fig. \ref{fig_Cl_diagonal_cov02} we show the measured errors, which are obtained as the square root of the diagonal elements of the numerical covariance matrix, $\hat{C}_{abij}$. These are compared with the theory, finding a generally good agreement, within 10\%. By looking at Fig. \ref{fig_Covariance_cl_pinocchio}, The fact that the numerical errors slightly underestimate the model predictions can be explained as follows. The partial sky coverage of the PLCs determines the emergence of a non-diagonal correlation, spread over an interval of multipoles given approximately by $ \pm 1/f_\mathrm{sky}$, as already found by \citet{Crocce11}. Since for our mocks $f_\mathrm{sky}=0.25$, the expected correlation width is $\Delta \ell=8$ and the measured non-diagonal terms rapidly drop to a noise consistent with zero. As discussed in \citet{Crocce11}, the integral of the elements of the covariance matrix arranged along a row gives us:
\begin{equation}
    \int \mathrm{d}\ell' \, \mathrm{Cov_{\ell \ell'}} \approx \frac{2}{(2\ell +1)\Delta \ell f_\mathrm{sky}}.
\end{equation}
In full-sky surveys, all the errors are summarised in the diagonal of the covariance matrix. However, the presence of boundaries causes a leakage of the diagonal errors to other $\ell$ modes, so that their values become lower than the simple $f_\mathrm{sky}$ boost given by Eq. \eqref{eq_covariance_Cl}. As we have seen in Fig. \ref{fig_Mixing_matrix_pinocchio}, thanks to the large sky coverage offered by the \textsc{Pinocchio} PLCs, almost all the correlation is concentrated within the band $\Delta \ell$. Therefore the diagonal model in Eq. \eqref{eq_covariance_Cl} is accurate enough to describe the numerical covariance. For smaller and highly fragmented regions of the sky, the approximation in Eq. \eqref{eq_covariance_Cl} does not hold. This is clearly visible in Fig. \ref{fig_Cl_diagonal_mixing_cov02}, where we show the result, after repeating the analysis but considering the 1004 deg$^2$ area subsample. Here, the diagonal approximation (black lines) overestimates the numerical errors in all redshift bins, with a more dramatic effect in the analysis of the reduced area. This issue can be resolved by relaxing the assumption of diagonality of the covariance matrix. In fact, we can think of the Kronecker delta as the full-sky limit of the mixing matrix, $R_{\ell\ell'}$. Thus, we can make the mixing matrix block-diagonal and compute the covariance as follows \citep{Loureiro19}:
\begin{equation}
    \mathrm{Cov_{\ell \ell'}}=\frac{2R_{\ell \ell'}}{(2\ell +1) f^2_\mathrm{sky}}\left(C_\ell+\frac{\Omega_\mathrm{sky}}{N_\mathrm{h}}\right)^2,
\end{equation}
with the second $f_\mathrm{sky}$ in the denominator coming up as the normalisation factor of the mixing matrix.
Furthermore, as we have already seen in Sect. \ref{Sect_angular_power_spectrum_map_pinocchio}, the role of the RSDs is only relevant at the largest angular scales and increases progressively with $z$. However, for $0.6<z<0.8$ and $0.8<z<1.0$, we find a percentage difference in errors between real and redshift space greater than 10\% only at $\ell<10$. To account also for nonlinearities at the BAO scale, we use the best-fit $\Sigma_\mathrm{NL}$ values of Eq. \eqref{eq_Pk_IR_resummation}, derived by fixing the cosmological parameters at \citet{Planck14}, the ones of the \textsc{Pinocchio} simulations. However, their role in the $C_\ell$ error estimate is completely negligible, so they are not considered in the plot. As noted by \citet{Crocce11}, the damping of the baryon acoustic features has no effect on the error predictions on the large angular scales. \\
\indent Analogously to what we found in Sect. \ref{Sect_angular_power_spectrum_map_pinocchio}, photo-$z$ uncertainties also determine a suppression in the error estimation, which is modelled by convolving the angular power spectrum model, $C_\ell$, with the Gaussian distribution in Eq. \eqref{eq_gaussian_perturbation_photoz}. 
\subsection{Covariance matrix of \texorpdfstring{$w(\theta)$}{} }
\label{sec_covariance_wtheta}
The angular correlation function can be derived analytically from $C_\ell$, by summation over an infinite set of multipoles, sampling progressively smaller angular scales:
\begin{equation}
    \label{eq_wtheta_from_cl}
     w(\theta)=\sum_{\ell} \left(\frac{2\ell+1}{4\pi}\right) L_\ell (\cos \theta) C_\ell,
\end{equation}
without the shot noise term, since it only contributes to the zero separation limit \citep{Chan18}, and also propagating systematic effects, such as RSDs and partial sky coverage. The covariance matrix of the angular correlation function is related to the $C_\ell$ one through \citep{Cabre07, Crocce11, Ross11}:
\begin{equation}
    \label{eq_covariance_wtheta}
    \mathrm{Cov}_{\theta_i \theta_j }=\sum_{\ell, \ell'}\left(\frac{2\ell+1}{4\pi}\right)^2 L_\ell (\mu_i) L_{\ell'} (\mu_j) \mathrm{Cov}_{\ell \ell'},  
\end{equation}
where $\theta_i$ and $\theta_j$ indicate the angular bins, while $\mu=\cos \theta$. This yields to: 
\begin{equation}
    \mathrm{Cov}_{\theta_i \theta_j }=\frac{2}{f_\mathrm{sky}} \sum_{\ell}\frac{2\ell+1}{(4\pi)^2} L_\ell (\mu_i) L_{\ell} (\mu_j) \left(C_\ell+\frac{1}{\bar{n}}\right)^2,    
\end{equation}
where we set the shot noise to $\Omega_\mathrm{sky}/N_\mathrm{h}=1/\bar{n}$ to simplify the notation. \\
\indent As for the power spectrum case, the $w(\theta)$ covariance matrix is modified by the fact that measurements are computed over finite angular width bins, which reduces the non-diagonal correlation. A correct estimate requires the bin-averaged Legendre polynomials, defined as \citep{Salazar14, Salazar17, Chan18}:
\begin{multline}
   \hat{L}_{\ell} \equiv \frac{\int_{\theta_{-}}^{\theta_{+}} L_{\ell}(\cos \theta) \sin \theta \mathrm{d} \theta}{\int_{\theta_{-}}^{\theta_{+}} \sin \theta \mathrm{d} \theta} = \\
    = \frac{L_{\ell+1}\left(\mu_{+}\right)-L_{\ell+1}\left(\mu_{-}\right)-L_{\ell-1}\left(\mu_{+}\right)+L_{\ell-1}\left(\mu_{-}\right)}{(2 \ell+1)\left(\mu_{-}-\mu_{+}\right)},
\end{multline}
where $\theta_+=\theta+\Delta \theta/2$ and $\theta_-=\theta-\Delta \theta/2$ represent the upper and the lower limits of the bin, while $\mu_+$ and $\mu_-$ are the corresponding cosines. This brings to \citep{Salazar14}: 
\begin{equation}
    \label{eq_covariance_w_salazar}
    \mathrm{Cov}_{\theta_i \theta_j }=\frac{2}{f_\mathrm{sky}} \sum_{\ell}\frac{2\ell+1}{(4\pi)^2} \hat{L}_\ell (\mu_i) \hat{L}_{\ell} (\mu_j) \left(C_\ell+\frac{1}{\bar{n}}\right)^2.    
\end{equation}
Notably, even the covariance matrix for partial sky coverage follows the inverse of $f_\mathrm{sky}$. However, this approximation might break when measuring $w(\theta)$ in highly fragmented survey areas, where the number of pairs as a function of separation does not simply scale with the effective area, $\Omega_\mathrm{sky}$, and thus it can not be reproduced by the shot noise term, $\Omega_\mathrm{sky}/N_\mathrm{h}$ \citep{Chan18}. \\
\indent Moreover, since the shot noise does not depend on the multipole $\ell$, we can make the binomial square explicit, and take $1/\bar{n}^2$ out from the summation. In particular, we can exploit the relation:
\begin{equation}
    \sum_\ell (2\ell+1)\hat{L}(\mu)\hat{L}(\mu')=\frac{2}{\mu_- - \mu_+} \delta_{K}^{\theta \theta'},
\end{equation}
obtaining \citep{Chan18}:
\begin{multline}
\label{eq_covariance_w_chan}
\mathrm{Cov}_{\theta_i \theta_j}=\frac{\delta_{K}^{\theta_i \theta_j   }}{4 \pi^2 f_{\mathrm{sky}} \bar{n}^2\left(\mu_{-}-\mu_{+}\right)}+\\+\sum_{\ell} \frac{2 \ell+1}{8 \pi^2 f_{\mathrm{sky}}} \hat{L}_{\ell}(\mu_i) \hat{L}_{\ell}\left(\mu_j \right)\left(C_{\ell}^2 + \frac{2C_{\ell} }{\bar{n}}\right),
\end{multline}
where the Kronecker $\delta_{K}^{\theta_i \theta_j}$ implies that the first term only acts on the diagonal of the covariance matrix.\\
\indent Actually, despite its mathematical equivalence, Eq. \eqref{eq_covariance_w_chan} is more convenient than Eq. \eqref{eq_covariance_w_salazar}. Indeed, the convergence to the final result is ensured by taking into account the contribution of infinite multipoles. Dealing with high-mass tracers, the shot noise term becomes dominant over the signal for $\ell \gtrsim 150$, depending on the redshift bin. From this scale, its contribution is dominant, while scale-dependent terms containing $C_\ell$ behave as secondary corrections. As a consequence, the analytical formulation of the first term of Eq. \eqref{eq_covariance_w_chan} allows the summation to be truncated to a reasonable $\ell_\mathrm{max}$, whereas with Eq. \eqref{eq_covariance_w_salazar} the absence of higher multipoles leads to an underestimation of the error at small $\theta$. 
\begin{figure*}[htbp]
\centering
\includegraphics[width=0.8\textwidth]{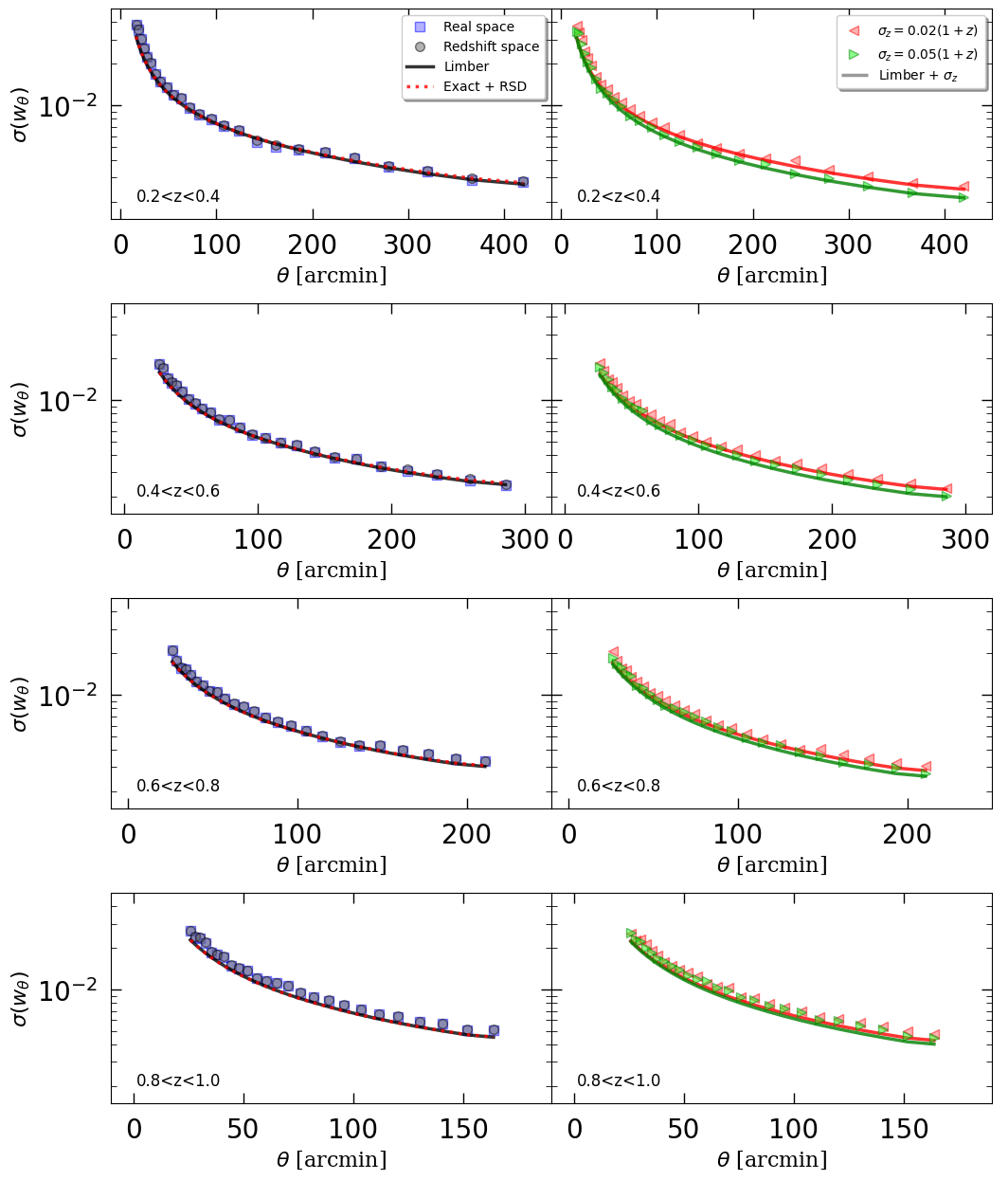}
\caption{Errors in the angular correlation function, derived as the square root of diagonal elements of $\mathrm{Cov}_{\theta_i \theta_j}$, in real (blue squares), redshift (with $\sigma_z=0$, black circles) and photo-$z$ space, with $\sigma_z=0.02$, (red triangles) and $\sigma_z=0.05$ (green triangles), from the numerical covariance matrix measured from 1000 mocks. The lines show the results obtained with the linear power spectrum in the Limber approximation (solid black line), with the exact integration given by
Eq. \eqref{eq_Cell_total}, which also includes the modelling of the RSDs (dashed red line) and with the damped
linear power spectrum in the Limber approximation (solid red and green lines).}
\label{fig_wtheta_diagonal_cov_02}
\end{figure*}
In Fig. \ref{fig_wtheta_diagonal_cov_02} we can see the errors in the angular correlation function, derived as the square root of the diagonal elements of $\mathrm{Cov}_{\theta_i \theta_j}$, in real, in redshift and in photo-$z$ space. As for $\mathrm{Cov}_{\ell \ell'}$, we notice that there is a good agreement between the numerical estimate and the theoretical model, with the former overestimating the latter by a maximum of 10\%, in the interval $0.8<z<1.0$. On the other hand, due to the range of angular scales involved, we see no differences between the Limber approximation and the $C_\ell$ model including the RSDs. This is valid also for the measurements, since the ratio between the real-space and redshift-space errors is always greater than $0.95$, meaning that there is no substantial difference between the two covariance matrices. In addition, we find that the theoretical red and green lines of Fig. \ref{fig_wtheta_diagonal_cov_02}, for $\sigma_{z,0}=0.02$ and $\sigma_{z,0}=0.05$ respectively, get progressively closer as the redshift increases. This is due to the fact that, at $z>0.6$, according to Eqs. \eqref{eq_covariance_Cl} and hence \eqref{eq_covariance_wtheta}, the errors becomes progressively more dominated by the shot noise and variations due to the $C_\ell$ term, given by photometric damping and RSDs, are no longer appreciable.\\
\begin{figure*}[htbp]
\centering
\includegraphics[width=1\textwidth]{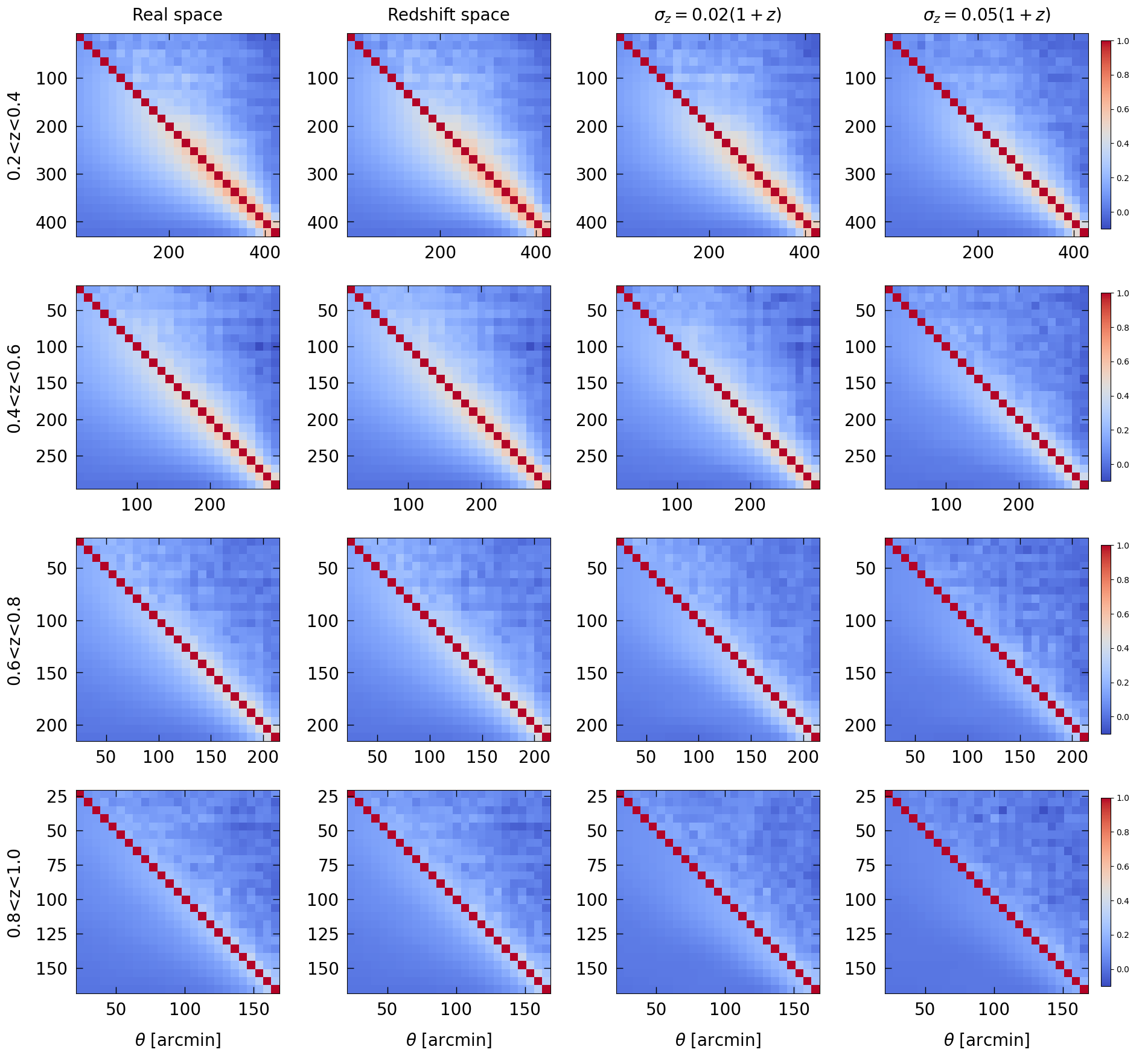}
\caption{Comparison between measurements and model in the full set of the $w(\theta)$ correlation matrices. The rows correspond to the different redshift bins, while different columns refer to real, redshift, with $\sigma_{z,0}=0$, and photo$-z$ space, with $\sigma_{z,0}=0.02$ and $\sigma_{z,0}=0.05$, from left to right. The upper triangles show the numerical estimate made on 1000 mocks. The lower triangles show the normalised covariance model.
}
\label{fig_wtheta_full_covariance_02}
\end{figure*}
\indent In Fig. \ref{fig_wtheta_full_covariance_02} we can see the full set of correlation matrices. The rows correspond to the different redshift bins and indicate that the correlation decreases with $z$, while the columns refer to the analysis in real, redshift or photo$-z$ space, with angular bins being less correlated in the presence of photometric errors. The upper triangles show the numerical estimate made on 1000 PLCs. The lower triangles show the bin-averaged model presented in Eq. \eqref{eq_covariance_w_chan}, where $C_\ell$s are computed with the Limber approximation.\\\\

\end{appendix}

\end{document}